\documentclass[sigconf,natbib=true]{acmart}
\renewcommand\footnotetextcopyrightpermission[1]{} 
\usepackage{graphicx}
\usepackage{amsmath}
\usepackage{booktabs}
\usepackage{algorithm}
\usepackage{algorithmic}
\usepackage{amsfonts}
\usepackage{multirow}[c]
\usepackage{makecell}
\usepackage{subfigure}
\usepackage{color}
\usepackage{bm}
\usepackage{epstopdf}
\usepackage{url}
\usepackage[cal=cm]{mathalfa}
\usepackage{balance}
\usepackage{threeparttable}
\usepackage{wrapfig}
\usepackage{tabularx}
\usepackage{array}
\usepackage{enumitem}
\usepackage{appendix}
\usepackage{tikz}

\setlist[itemize]{leftmargin=*}

\makeatletter
\newcommand\notsotiny{\@setfontsize\notsotiny\@vipt\@viipt}
\makeatother

\AtBeginDocument{%
  \providecommand\BibTeX{{%
    \normalfont B\kern-0.5em{\scshape i\kern-0.25em b}\kern-0.8em\TeX}}}

\begin{document}
\author{Xiaodong Li\textsuperscript{1,3}, Ruochen Yang\textsuperscript{1,3}, Shuang Wen\textsuperscript{2}, Shen Wang\textsuperscript{2}, Yueyang Liu\textsuperscript{2}, Guoquan Wang\textsuperscript{2}, Weisong Hu\textsuperscript{2}, Qiang Luo\textsuperscript{2}, Jiawei Sheng\textsuperscript{1}, Tingwen Liu\textsuperscript{1,3}, Jiangxia Cao\textsuperscript{2}*, Shuang Yang\textsuperscript{2}, Zhaojie Liu\textsuperscript{2}}
\thanks{* Project leader. This work was funded by Kuaishou.}
\affiliation{
\institution{\textsuperscript{1}Institute of Information Engineering, Chinese Academy of Sciences, Beijing, China\\
\textsuperscript{2}Kuaishou Technology, Beijing, China\\
\textsuperscript{3}School of Cyber Security, University of Chinese Academy of Sciences, Beijing, China}
\country{\{lixiaodong, yangruochen, shengjiawei, liutingwen\}@iie.ac.cn, \{wenshuang, wangshen, liuyueyang05,\\ wangguoquan03, huweisong, luoqiang, caojiangxia, yangshuang08, zhaotianxing\}@kuaishou.com}
}
\renewcommand{\shortauthors}{Xiaodong Li et al.}

\title{FARM: Frequency-Aware Model for \\ Cross-Domain Live-Streaming Recommendation}

\setcopyright{acmcopyright}
\copyrightyear{2025}
\acmYear{2025}
\acmDOI{}

\acmConference[Conference’17]{}{July 2017}{Washington, DC, USA}

\renewcommand{\shorttitle}{FARM: Frequency-Aware Model for Cross-Domain Live-Streaming Recommendation}

\begin{abstract}
Live-streaming services have attracted widespread popularity due to their real-time interactivity and entertainment value. Users can engage with live-streaming authors by participating in live chats, posting likes, or sending virtual gifts to convey their preferences and support.
However, the live-streaming services faces serious data-sparsity problem, which can be attributed to the following two points:
(1) User's valuable behaviors are usually sparse, \textit{e.g.}, \textit{like}, \textit{comment} and \textit{gift}, which are easily overlooked by the model, making it difficult to describe user's personalized preference. (2) The main exposure content on our platform is short-video, which is 9 times higher than the exposed live-streaming, leading to the inability of live-streaming content to fully model user preference.
To this end, we propose a \textit{\textbf{F}requency-\textbf{A}ware \textbf{M}odel for Cross-Domain Live-Streaming \textbf{R}ecommendation}, termed as \textbf{FARM}. Specifically, we first present the intra-domain frequency aware module to enable our model to perceive user's sparse yet valuable behaviors, \textit{i.e.}, high-frequency information, supported by the Discrete Fourier Transform (DFT). To transfer user preference across the short-video and live-streaming domains, we propose a novel preference align before fuse strategy, which consists of two parts: the cross-domain preference align module to align user preference in both domains with contrastive learning, and the cross-domain preference fuse module to further fuse user preference in both domains using a serious of tailor-designed attention mechanisms.
Extensive offline experiments and online A/B testing on Kuaishou live-streaming services demonstrate the effectiveness and superiority of FARM. Our FARM has been deployed in online live-streaming services and currently serves hundreds of millions of users on Kuaishou.

\end{abstract}



\maketitle

\begin{figure}[t]
\setlength{\abovecaptionskip}{0.cm}
	\begin{center}
        \subfigure
        {\begin{minipage}[b]{.47\linewidth}
        \centering
        \includegraphics[scale=0.31]{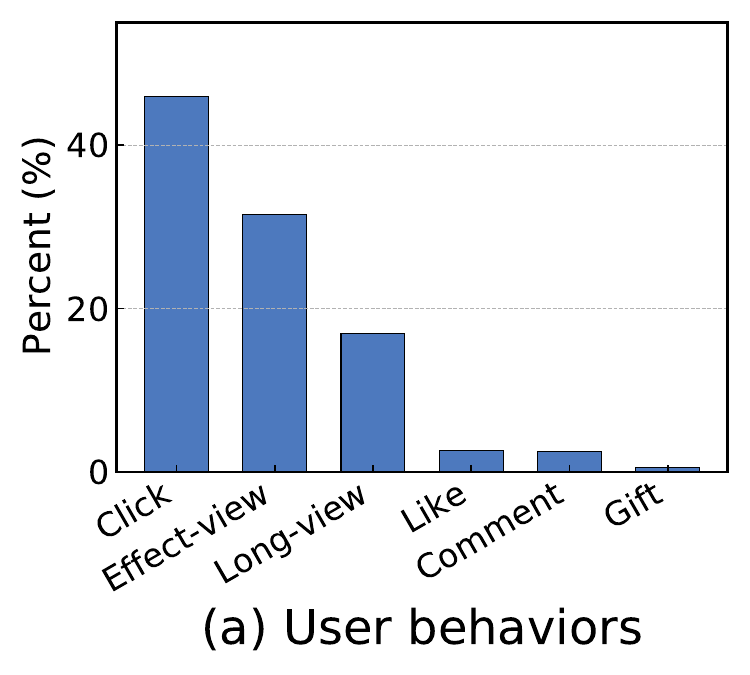}
        \end{minipage}}
        \subfigure
        {\begin{minipage}[b]{.47\linewidth}
        \centering
        \includegraphics[scale=0.31]{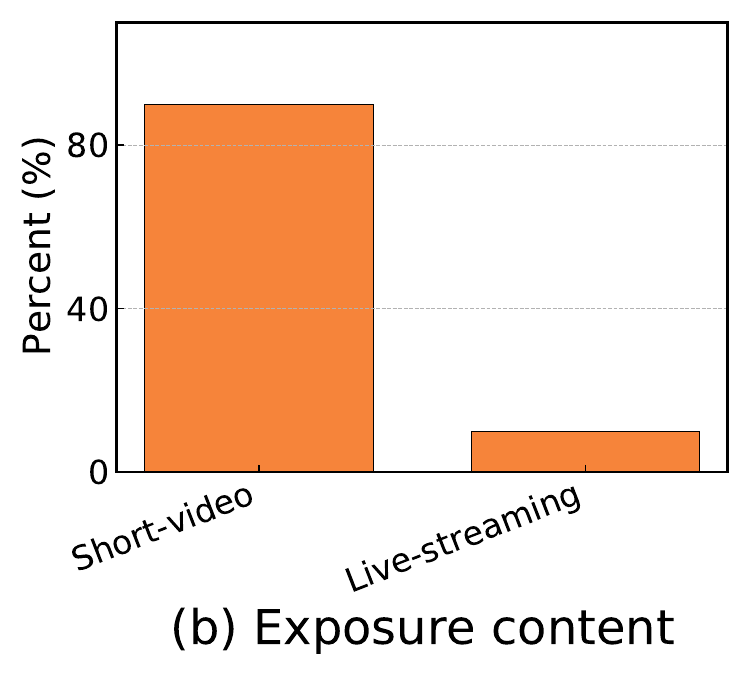}
        \end{minipage}}
        
	\caption{An illustration of (a) relative percentages of different user behaviors; (b) relative percentages of exposure content in our live-streaming services on Kuaishou.}
	\label{motivation}
	\end{center}
\vspace{-0.9em}
\end{figure}

\section{Introduction}
Short-video and Live-streaming applications like Kuaishou and TikTok have grown rapidly in recent years. A large number of active users watch live-streamings every day, and express their preference and support to the live-streaming authors by posting comments, giving likes or virtual gifts. Consequently, the personalized recommendation systems~\cite{twin,dien} is the foundation of our live-streaming services, which play a crucial role in accurately recommending appropriate authors to enhance user's experience.

Recently, several studies~\cite{mmbee,kuaihl,contentctr} have been proposed to improve the performance of live-streaming recommendation. Specifically, ContentCTR~\cite{contentctr} and KuaiHL~\cite{kuaihl} propose the multi-modal transformer that can make full use of complex multi-modal interactions to identify the attractive moment of live-streaming authors. MMBee~\cite{mmbee} constructs metapaths to capture multi-hop relationships between users and live-streaming authors, thus enriching the user's historical behavior sequences. 
More recently, some studies~\cite{eliverec,lcn} further consider cross-domain recommendation (CDR)~\cite{cdrnp,dmcdr} to improve the recommendation performance in the live-streaming domain via the preference transfer from another relevant domain.
For instance, eLiveRec~\cite{eliverec} employs the disentangled encoder to learn user's domain-shared and domain-specific intensions across domains.
LCN~\cite{lcn} introduces a lifelong attention pyramid with three cascading attention levels to mitigate the impact of the noise information in cross-domain sequences.

We perform a statistical analysis of the live-streaming services on our platform and find that it faces serious \textbf{data-sparsity} issue, which limits the capability of our live-streaming model. Specifically, from the statistical results shown in Figure~\ref{motivation}, we discover two phenomena that lead to the data-sparsity issue:
(1) \textbf{Sparsity of valuable user behaviors:} User may leave multiple behaviors while watching the live-streamings, \textit{e.g.}, \textit{click}, \textit{long-view}, \textit{gift}, etc. However, user's valuable behaviors (\textit{e.g.}, \textit{like}, \textit{comment} and \textit{gift}) are usually sparse, which are easily overlooked by our model. To better understand this phenomenon, we give the relative percentages of different user behaviors in our live-streaming services in Figure~\ref{motivation}(a). We observe that compared with behaviors such as \textit{click}, \textit{effect-view} and \textit{long-view}, behaviors such as \textit{like}, \textit{comment} and \textit{gift} are extremely sparse, leading to the challenge of accurately describing user's personalized preference. (2) \textbf{Sparsity of exposure content:} The main exposure content on our Kuaishou platform is short-video, which is 9 times higher than the exposed live-streaming (as shown in Figure~\ref{motivation}(b)), resulting in the user's short-video viewing time being much longer than that of live-streaming. Therefore, users may leave more behaviors with the short-video content, while the live-streaming content alone cannot fully model user preference.
However, previous cross-domain live-streaming recommendation methods~\cite{eliverec,lcn} lack tailor-designed modeling to address the live-streaming specific data-sparsity issues as mentioned above.

To address the above data-sparsity issues, this paper proposes \textbf{FARM}, a novel \textit{\textbf{F}requency-\textbf{A}ware \textbf{M}odel for Cross-Domain Live-Streaming \textbf{R}ecommendation}. Specifically, to alleviate the first challenge, we present the intra-domain frequency-aware module to enable our model to perceive user's sparse yet valuable behavior (\textit{i.e.}, high-frequency information) in both the short-video and live-streaming domains, in accordance with the Discrete Fourier Transform (DFT) theory. In addition, aiming to address the second challenge, we propose a novel preference align before fuse strategy, which consists of two parts: the cross-domain preference align module applies contrastive learning to align user preference in both domains, which narrows the gap between user preference representations in both domains. Meanwhile, the cross-domain preference fuse module adopts a serious of tailor-designed attention mechanisms to further fuse user preference in both domains, thus achieving the transfer of user preference across the short-video and the live-streaming domains.

Our main contribution can be summarized as follows:
\begin{itemize}
    \item We reveal the data-sparsity problem in our live-streaming services on Kuaishou, including the sparsity of valuable user behaviors, and the sparsity of exposure content.
    \item We propose a novel Frequency-Aware Model for Cross-Domain Live-Streaming Recommendation, termed as FARM, which perceive user's sparse yet valuable behaviors using the intra-domain frequency-aware module, and further transfer user preference across domains with the preference align before fuse strategy.
    \item We conduct extensive offline experiments in live-streaming services on Kuaishou to validate the effectiveness of FARM. Online A/B testing further demonstrate that FARM brings considerable revenue growth (up to 0.41\% improvement in Click) to the platform. FARM has been deployed in online live-streaming services and currently serves hundreds of millions of users on Kuaishou.
\end{itemize}

\section{Preliminary}
This section introduces the background knowledge of the frequency domain, the Fourier Transform and explains the problem definition for cross-domain live-streaming recommendation.

\subsection{Fourier Transform}\label{FFT}
Discrete Fourier Transform (DFT) has been widely explored in diverse fields including graph signal processing~\cite{sgnn,fgnn,cdgsp} and recommendation systems~\cite{gf-cf,higsp,pgsp,bsarec}.
Typically, DFT is a transformation used to convert discrete time domain signals into discrete frequency domain (Fourier domain) representations, which can be denoted as $\mathcal{F}:\mathbb{R}^N\to\mathbb{C}^N$. While the Inverse Discrete Fourier Transform (IDFT) can be represented as $\mathcal{F}^{-1}:\mathbb{C}^N\to\mathbb{R}^N$, aiming to transform the discrete frequency domain representations back into discrete time domain signals.

Specifically, applying $\mathcal{F}$ to a signal can be interpreted as performing a DFT matrix, which can be defined as: 
\begin{equation}
\begin{split}
\bm{F}=[\bm{f}_1,\bm{f}_2,\dots,\bm{f}_N]/\sqrt{N},
\end{split}
\label{}
\end{equation}
The rows of this matrix are composed of the Fourier basis, which can be formulated as:
\begin{equation}
\begin{split}
\bm{f}_j=[e^{2\pi i(j-1)\cdot 0}\dots e^{2\pi i(j-1)(N-1)}]^\top/\sqrt{N},
\end{split}
\label{}
\end{equation}
where $\bm{f}_j\in \mathbb{R}^N$, $i$ is the imaginary unit and $j$ denotes the $j$-th row. Furthermore, the IDFT can be expressed as:
\begin{equation}
\begin{split}
\bm{F}^{-1}=\bm{F}^H/\sqrt{N},
\end{split}
\label{}
\end{equation}
where $\bm{F}^H$ represents the conjugate transpose of $\bm{F}$. The normalization factor $1/\sqrt{N}$ ensures the unitary nature of the transformation.

Given $\bm{x}$ in discrete time domain, we represent the spectrum of $\bm{x}$ as $\bar{\bm{x}}=\mathcal{F}\bm{x}$. Consequently, we can define $\bar{\bm{x}}_L$ with $c$ lowest elements of $\bar{\bm{x}}\in \mathbb{C}^c$, and the remaining elements can be represented as $\bar{\bm{x}}_H\in \mathbb{C}^{N-c}$.
Formally, the low-frequency components of the sequential signal $\bm{x}$ can be defined as follows:
\begin{equation}
\begin{split}
\text{Low}[\bm{x}]=[\bm{f}_1,\bm{f}_2,\dots,\bm{f}_c]\bar{\bm{x}}_{L}\in \mathbb{R}^N,
\end{split}
\label{}
\end{equation}
Similarly, the high-frequency components can be defined as follows:
\begin{equation}
\begin{split}
\text{High}[\bm{x}]=[\bm{f}_{c+1},\bm{f}_{c+2},\dots,\bm{f}_N]\bar{\bm{x}}_{H}\in \mathbb{R}^N,
\end{split}
\label{}
\end{equation}

In our FARM, the low-frequency information reflects user's dense behaviors (\textit{e.g.}, click and long-view), while the high-frequency information represents user's sparse behaviors (\textit{e.g.}, like, comment and gift). Our goal is to enable model to perceive sparse yet valuable high frequency information.

\subsection{Problem Definition}

In this paper, we conceptualize the cross-domain live-streaming recommendation as leveraging date from the short-video domain to improve the recommendation performance in the live-streaming domain.
Specifically, in industry, we focus on predicting multiple probabilities that reflect a user's preference for a candidate live-streaming author in multiple aspects, including:
the \textit{click} probability (\textit{i.e.}, \textbf{CTR}), the \textit{effective-view} probability (\textit{i.e.}, \textbf{ETR}), the \textit{long-view} probability (\textit{i.e.}, \textbf{LVTR}), the \textit{like} probability (\textit{i.e.}, \textbf{LTR}), the \textit{comment} probability (\textit{i.e.}, \textbf{CMTR}), and the \textit{gift} probability (\textit{i.e.}, \textbf{GTR}). 

Formally, let $\bm{x}^{\text{candidate}}_l$ denote the candidate live-streaming features (\textit{e.g.}, item tags, live ID, author ID, user ID, etc.), and $\bm{X}_s/\bm{X}_l$ denote the user's historical interaction log sequence in short-video/live-streaming domains, where $\bm{x}_s\in\bm{X}_s$, $\bm{x}_l\in\bm{X}_l$ represents multiple distinct categories of features, including ID-based features and Side Info-based features as follows:
\begin{itemize}
    \item ID-based features refer to the author (item) ID of user's historical viewing, denoted as $\bm{x}^{\mathrm{Aid}}$.
    \item Side Info-based features refer to the author's attribute information (\textit{e.g.}, $\bm{x}^{\mathrm{Page}}$, $\bm{x}^{\mathrm{Tag}}$, $\bm{x}^{\mathrm{Cluster}}$) and the user's behavior information towards the author (\textit{e.g.}, $\bm{x}^{\mathrm{Play}}$, $\bm{x}^{\mathrm{Lag}}$, $\bm{x}^{\mathrm{Label}}$).
\end{itemize} 
Thus, given samples consisting of ID-based features, Side Info-based features from both domains, and several true labels (\textit{e.g.}, $y^{ctr}\in\{0,1\}$, $y^{etr}\in\{0,1\}$, $y^{lvtr}\in\{0,1\}$, etc.) in live-streaming domain, the model aims to predict the probability of \textbf{XTRs} (\textit{e.g.}, CTR, ETR, LVTR, etc.) for user with respect to the candidate author in the live-streaming domain as follows:
\begin{equation}
\begin{split}
\hat{y}^{ctr},\hat{y}^{etr},\dots,\hat{y}^{gtr}=f_{\theta}\big([\bm{X}_s, \bm{X}_l], \bm{x}^{\text{candidate}}_l\big),
\end{split}
\label{}
\end{equation}
where $\hat{y}^{ctr},\hat{y}^{etr},\dots,\hat{y}^{gtr}$ are predicted probabilities, $f_{\theta}$ is a multi-task module which can be implemented by MMoE~\cite{mmoe} and PLE~\cite{ple}.

Ultimately, the multi-task module $f_\theta$ is optimized by minimizing the standard negative log-likelihood function, which can be formulated as follows:
\begin{equation}
\begin{split}
\mathcal{L}_{XTRs}=-\sum^{ctr,\dots,gtr}_{xtr}(y^{xtr}\mathrm{log}(\hat{y}^{xtr})+(1-y^{xtr})\mathrm{log}(1-\hat{y}^{xtr})),
\end{split}
\label{loss_xtrs}
\end{equation}
where $y^{xtr}$ represents the true labels of user feedback, and $\hat{y}^{xtr}$ is the predicted XTRs.
In the following section, we will express how our FARM perceives user's sparse yet valuable behaviors and further fuses user's short-video and live-streaming preferences to serve our live-streaming services on Kuaishou.

\begin{figure*}[t!]
\begin{center}
\includegraphics[width=18cm]{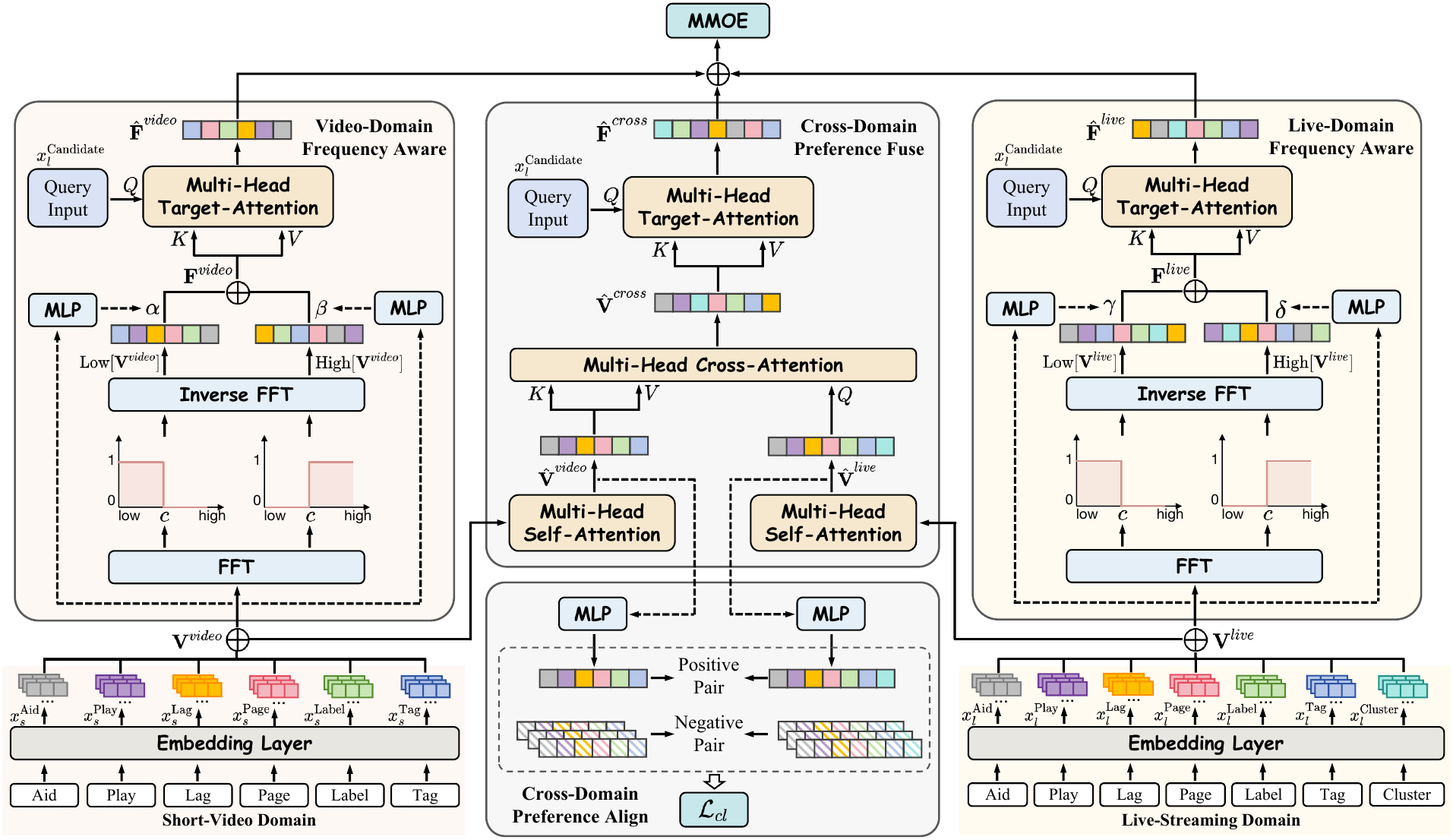}
\caption{The overall framework of FARM. The intra-domain frequency-aware module enable the model to perceive user's sparse yet valuable behaviors in both domains. The preference align before fuse strategy first apply the contrastive learning to align user preference in both domains, and further adopt a serious of tailor-designed attention mechanisms to fuse user preference in both domains, thus achieving the preference transfer across the short-video and live-streaming domains.}
\label{main_model}
\end{center}
\end{figure*}

\section{Methodology}
In this section, we introduce FARM, a frequency aware model for cross-domain live-streaming recommendation, as described in Figure~\ref{main_model}. Our FARM mainly consists of four parts: 1) the \textbf{embedding layer} initializes embeddings of user's historical behavior sequence in both domains, 2) the \textbf{intra-domain frequency-aware module} aims to enable our model to perceive sparse yet valuable high-frequency information, 
3) the \textbf{cross-domain preference align module} adopts contrastive learning to ensure user preference in the live-streaming domain align closely with user preference in the short-video domain,
and 4) the \textbf{cross-domain preference fuse module} applies a series of well-designed attention mechanisms to fuse user preference in both domains for better recommendation in the live-streaming domain.
Details of each module are demonstrated in the following parts.

\subsection{Embedding Layer}
The embedding layer provides the initialized embeddings of user's historical behavior sequence in both domains. Specifically, we convert the user's ID-based features and Side Info-based features into a unified low-dimension latent space, which can be formulated as:
\begin{equation}
\begin{split}
\mathbf{V}^{video}&=\text{Concat}\big(\big[\bm{x}^{\mathrm{Aid}}_s,\bm{x}^{\mathrm{Play}}_s,\dots,\bm{x}^{\mathrm{Tag}}_s\big]\big),\\
\mathbf{V}^{live}&=\text{Concat}\big(\big[\bm{x}^{\mathrm{Aid}}_l,\bm{x}^{\mathrm{Play}}_l,\dots,\bm{x}^{\mathrm{Tag}}_l,\bm{x}^{\mathrm{Cluster}}_l\big]\big),
\end{split}
\label{}
\end{equation}
where \textit{Concat} represents the concatenation operation. $\mathbf{V}^{video}\in \mathbb{R}^{L\times D_1}$ and $\mathbf{V}^{live}\in \mathbb{R}^{L\times D_2}$ denote embeddings of user's historical behavior sequence in short-video domain and live-streaming domain, respectively, where $L$ is the sequence length, $D_1$ and $D_2$ are the embedding dimension.

\subsection{Intra-Domain Frequency-Aware Module}\label{fq_modeling}
In real-world industrial environment (\textit{e.g.}, live-streaming domain), user's valuable behaviors (\textit{e.g.}, like, comment and gift) could accurately describe user's personalized preference. However, these valuable behaviors are usually sparse, which are easily overlooked by the model. To this end, we introduce the intra-domain frequency-aware module to enable our model to perceive user's sparse yet valuable behaviors (\textit{i.e.}, high-frequency information), supported by Discrete Fourier Transform (DFT)~\cite{fgnn,bsarec} reviewed in Section~\ref{FFT}.

Specifically, the user's historical behavior sequence can be divided into low-frequency and high-frequency components:
\begin{equation}
\begin{split}
\mathbf{F}^{video}&=\alpha\text{Low}[\mathbf{V}^{video}]+\beta\text{High}[\mathbf{V}^{video}],\\
\mathbf{F}^{live}&=\gamma\text{Low}[\mathbf{V}^{live}]+\delta\text{High}[\mathbf{V}^{live}],
\end{split}
\label{eq_fft}
\end{equation}
where $\mathbf{F}^{video}$ and $\mathbf{F}^{live}$ contain user's valuable behaviors in the short-video domain and live-streaming domain, respectively. 
Besides, $\alpha$ and $\gamma$ are the trainable parameters to control the strength of low-pass components, while $\gamma$ and $\delta$ control the strength of high-pass components, which can be defined as follows:
\begin{equation}
\begin{split}
\alpha&=\text{Sigmoid}(\text{MLP}(\mathbf{V}^{video})),
~\beta=\text{Sigmoid}(\text{MLP}(\mathbf{V}^{video})),\\
\gamma&=\text{Sigmoid}(\text{MLP}(\mathbf{V}^{live})),
\quad\delta=\text{Sigmoid}(\text{MLP}(\mathbf{V}^{live})),
\end{split}
\label{eq_trainable}
\end{equation}
where we use a sigmoid activation function to obtain weights (\textit{e.g.}, $\alpha$, $\beta$, $\gamma$ and $\delta$) ranging from 0 to 1.

Ultimately, we utilize the candidate live-streaming features (\textit{i.e.}, $\bm{x}^{\text{candidate}}_l$) as a query and employ a multi-head target-attention mechanism~\cite{trans} to enable our model to aware the high-frequency components from various perspectives, which can be written as:
\begin{equation}
\begin{split}
\hat{\mathbf{F}}^{video}&=\text{MH-Target-Att}(Q=\bm{x}^{\text{candidate}}_l,K=\mathbf{F}^{video},V=\mathbf{F}^{video}),\\
\hat{\mathbf{F}}^{live}&=\text{MH-Target-Att}(Q=\bm{x}^{\text{candidate}}_l,K=\mathbf{F}^{live},V=\mathbf{F}^{live}),
\end{split}
\label{frequency_multi_head}
\end{equation}
where $\hat{\mathbf{F}}^{video}$ and $\hat{\mathbf{F}}^{live}$ are the final inputs to MMOE.

\subsection{Preference Align before Fuse Strategy}
On the slide page of our industrial deployment, due to uneven traffic distribution, 90\% of the exposure content is short-video and 10\% is live-streaming (refer to Figure~\ref{motivation}(b)), which leads to the data sparsity issue in the live-streaming domain. 
Consequently, how to transfer user preference in the short-video domain to the live-streaming domain is the core task to improve the recommendation performance in the live-streaming domain.

To this end, we first introduce a cross-domain preference align module to align user preference in both domains using contrastive learning, which narrows the gap between user preference representations in both domains. Furthermore, we design a cross-domain preference fuse module, which utilizes a serious of tailor-designed attention mechanisms~\cite{trans} to fuse user preference in both domains, thus achieving the transfer of user preference across domains.

\subsubsection{\textbf{Cross-Domain Preference Align Module}}\label{cl_modeling}
Inspired by the success of previous works~\cite{semi,sitn,abf}, the use of contrastive learning to align the user preference in both domains before fusing them is effective and critical.
Specifically, user preferences in the short-video domain and live-streaming domain are self-decoupled by multi-head self-attention as follows: 
\begin{equation}
\begin{split}
\hat{\mathbf{V}}^{video}&=\text{MH-Self-Att}(Q=\mathbf{V}^{video}, K=\mathbf{V}^{video},V=\mathbf{V}^{video}),\\
\hat{\mathbf{V}}^{live}&=\text{MH-Self-Att}(Q=\mathbf{V}^{live}, K=\mathbf{V}^{live},V=\mathbf{V}^{live}),
\end{split}
\label{self_att}
\end{equation}
where $\hat{\mathbf{V}}^{video}$ and $\hat{\mathbf{V}}^{live}$ represent user preference in both domains. We then leverage Multi-Layer Perceptron (\textit{i.e.}, MLP) to refine user preference in both domains:
\begin{equation}
\begin{split}
\bm{h}^{video} &= \text{MLP}(\texttt{Mean}(\hat{\mathbf{V}}^{video})),\\
\bm{h}^{live} &= \text{MLP}(\texttt{Mean}(\hat{\mathbf{V}}^{live})),
\end{split}
\label{eq_mlp}
\end{equation}
where the $\texttt{Mean}(\cdot): \mathbb{R}^{L\times D}\to \mathbb{R}^{D}$ is a simple pooling function to compress sequence representation, $\bm{h}^{video}$ and $\bm{h}^{live}$ are refined user preference in both domains. 
Afterward, we construct positive and negative sample pairs for contrastive learning, which is grounded in the understanding that user preferences tend to be similar across different domains. This similarity enables us to uniformly sample both positive and negative pairs across domains as follows:
\begin{itemize}
    \item Positive sample pair: The user preference $\bm{h}^{video}$ in the short-video domain and the preference $\bm{h}^{live}$ in the live-streaming domain form a positive pair. 
    \item Negative sample pair: Since our goal is to make recommendation in live-streaming domain, we perform contrastive learning from video to live. Specifically, the user preference representation in live-streaming domain is fixed, while that in short-video domain is iteratively sampled from different users within the same batch.
\end{itemize}
After obtaining the positive and negative sample pairs, we perform contrastive loss function to enforce the model to narrow the gap between user preference representations for positive pairs within a given batch as follows:
\begin{equation}
\begin{split}
\mathcal{L}_{cl}=-\frac{1}{|B|}\sum_{i=1}^{B}\log\frac{\exp\left(\text{sim}(\bm{h}^{live}_i,\bm{h}^{video}_i)/\tau\right)}{\sum_{j=1}^{B} \exp\left(\text{sim}(\bm{h}^{live}_i,\bm{h}^{video}_j)/\tau\right)},
\end{split}
\label{loss_cl}
\end{equation}
where $B$ denotes the size of training batch, $\text{sim}(\cdot)$ is a dot product, and $\tau$ represents the temperature coefficient.

The overall loss function of our FARM is the combination of the $\mathcal{L}_{XTRs}$ in Eq.~(\ref{loss_xtrs}) and the contrastive loss in Eq.~(\ref{loss_cl}):
\begin{equation}
\begin{split}
\mathcal{L}=\mathcal{L}_{XTRs}+\lambda\mathcal{L}_{cl},
\end{split}
\label{overall_loss}
\end{equation}
where $\lambda$ is a hyper-parameter that controls the strength of contrastive loss, ranging from 0 to 1.

\subsubsection{\textbf{Cross-Domain Preference Fuse Module}}\label{fuse_modeling}
To achieve the preference transfer across domains, we design a serious of customized attention mechanisms to fuse user preference in both the short-video and live-streaming domains. 

We start by the multi-head self-attention to obtain user preference representation $\hat{\mathbf{V}}^{video}$ in the short-video domain and $\hat{\mathbf{V}}^{live}$ in the live-streaming domain (refer to Eq.~(\ref{self_att})), respectively. Since the contrastive loss has been designed to align user preference in both domains, we could further adopt multi-head cross-attention to fuse user preference in both domains. Specifically, as our goal is to improve the recommendation performance in live-streaming domain, we leverage user preference representation in the live-streaming domain (\textit{i.e.}, $\hat{\mathbf{V}}^{live}$) as a query, which could explicitly transfer beneficial user preference existing in the short-video domain:  
\begin{equation}
\begin{split}
\hat{\mathbf{V}}^{cross}=\text{MH-Cross-Att}(Q=\hat{\mathbf{V}}^{live}, K=\hat{\mathbf{V}}^{video},V=\hat{\mathbf{V}}^{video}),
\end{split}
\label{eq_cross_attention}
\end{equation}
where $\hat{\mathbf{V}}^{cross}$ represents the fused user preference in both domains. Afterward, similar to Eq.~(\ref{frequency_multi_head}), we utilize the candidate live-streaming features (\textit{i.e.}, $\bm{x}^{\text{candidate}}_l$) as a query and apply a multi-head target-attention to achieve a fine-grained preference extraction:
\begin{equation}
\begin{split}
\hat{\mathbf{F}}^{cross}=\text{MH-Target-Att}(Q=\bm{x}^{\text{candidate}}_l,K=\hat{\mathbf{V}}^{cross},V=\hat{\mathbf{V}}^{cross}),
\end{split}
\label{fuse_multi_head}
\end{equation}
where $\hat{\mathbf{F}}^{cross}$ is the final input to MMOE. In summary, the overall inputs to MMOE can be formulated as follows:
\begin{equation}
\begin{split}
\hat{\mathbf{F}}=\text{Concat}\big(\big[\hat{\mathbf{F}}^{video},\hat{\mathbf{F}}^{live},\hat{\mathbf{F}}^{cross}\big]\big),
\end{split}
\label{overall_inputs}
\end{equation}
where $Concat$ represents the concatenation operation.

\section{Experiments}
In this section, we conduct extensive offline experiments and online A/B testing on our Kuaishou live-streaming services (\textit{i.e.}, Kuaishou and Kuaishou Lite) to answer the following research questions:
\begin{itemize}
    \item \textbf{RQ1:} How do different modules of our FARM perform in offline experiments?
    \item \textbf{RQ2:} Does the intra-domain frequency-aware module proposed in FARM really perceive user's sparse yet valuable behaviors?
    \item \textbf{RQ3:} Does the cross-domain preference align module proposed in FARM really narrow the gap between user preference representation in both domains?
    \item \textbf{RQ4:} Does the cross-domain preference fuse module proposed in FARM really achieve the preference transfer across domains?
    \item \textbf{RQ5:} Does our FARM encourage more low-active (sparse behaviors) users to watch live-streamings?
    \item \textbf{RQ6:} How does our FARM perform in online A/B testing?
\end{itemize}

\subsection{Experimental Settings}
\subsubsection{\textbf{Kuaishou Dataset}} 
We conduct offline experiments and online A/B testing on Kuaishou live-streaming services, including \textbf{Kuaishou} and \textbf{Kuaishou Lite}. In this paper, we focus on the short-video domain and live-streaming domain. Specifically, the live-streaming domain includes about 3 billion interaction logs with live-streaming content in Kuaishou application, which are collected by the \textbf{real-time reporter with first-only strategy}~\cite{moment_cross}: We apply a 30s sliding window to generate the data-streaming segment samples. The reporting mechanism is that, the positive labels are reported immediately, and we only learn the first positive occurrence for each behavior, while other samples will be ignored. On the contrary, 
the negative labels are reported when user exits the live-streaming.
In industry deployment, different businesses are employed separately. For instance, the user's short-video interaction logs are collected by the short-video data-streaming engine, which can only be consumed by the short-video model for recommendation. Thus, we cannot directly consume short-video data-streaming, and our live-streaming model can only be supervised by the live-streaming data-streaming. Fortunately, we have established historical storage services to preserve user's interaction logs~\cite{sdim,eta}. Additionally, the data-streaming engine can obtain user's interaction logs from other business systems upon request, which are integrated as part of the input features. Consequently, we could obtain user's short-video interaction logs to make cross-domain live-streaming recommendation.

The Kuaishou dataset consists of two components: the training set and the testing set. Specifically, the training set comprises users' real interaction logs collected over a seven-day period, encompassing all short-video and live-streaming content during that time to train our model. The testing set is sampled from user interaction logs recorded on the subsequent day following the training set collection period and is utilized to evaluate the model's performance.

\begin{table*}[t]
\small
\centering
\caption{Offline performance(\%) of FARM in terms of AUC, UAUC and GAUC in live-streaming services on Kuaishou. Improve $\uparrow$ represents the improvements(\%) of FARM over PLE (base) model.}
\label{mainexperiment_1}
\setlength\tabcolsep{2.0pt}
\begin{tabular*}{1 \textwidth}
{@{\extracolsep{\fill}}@{}lcccccccccccccccccc@{}}
\toprule
\multirow{2.5}{*}{\textbf{Variants}} & \multicolumn{3}{c}{\textbf{Click}}  & \multicolumn{3}{c}{\textbf{Effective-view}} & \multicolumn{3}{c}{\textbf{Long-view}} & \multicolumn{3}{c}{\textbf{Like}} & \multicolumn{3}{c}{\textbf{Comment}} & \multicolumn{3}{c}{\textbf{Gift}}
\\ \cmidrule(r){2-4} \cmidrule(r){5-7} \cmidrule(r){8-10} \cmidrule(r){11-13} \cmidrule(r){14-16} \cmidrule(r){17-19}& AUC & UAUC & GAUC & AUC & UAUC & GAUC & AUC & UAUC & GAUC & AUC & UAUC & GAUC & AUC & UAUC & GAUC & AUC & UAUC & GAUC\\
\midrule
PLE (Base) & 81.90 & 60.55 & 60.52 & 80.78 & 63.21 & 63.11 & 87.68 & 71.97 & 72.22 & 91.79 & 67.31 & 67.19 & 93.68 & 69.37 & 69.05 & 96.27 & 66.39 & 65.82 \\
\midrule
FARM & 82.35 & 61.18 & 61.16 & 80.96 & 63.50 & 63.42 & 88.27 & 73.72 & 73.93 & 92.29 & 67.77 & 67.52 & 94.18 & 70.78 & 70.52 & 96.61 & 68.92 & 68.46       \\
Improve $\uparrow$ & \textbf{+0.45} & \textbf{+0.63} & \textbf{+0.64} & \textbf{+0.18} & \textbf{+0.29} & \textbf{+0.31} & \textbf{+0.59} & \textbf{+1.75} & \textbf{+1.71} & \textbf{+0.50} & \textbf{+0.46} & \textbf{+0.33} & \textbf{+0.50} & \textbf{+1.41} & \textbf{+1.47} & \textbf{+0.34} & \textbf{+2.53} & \textbf{+2.64}   \\
\midrule
\textit{w}/\textit{o} V-FA & 82.33 & 61.12 & 61.11 & 80.93 & 63.49 & 63.40 & 88.25 & 73.71 & 73.91 & 92.23 & 67.52 & 67.39 & 94.13 & 70.64 & 70.37 & 96.54 & 68.11 & 68.15      \\
\textit{w}/\textit{o} L-FA & 82.23 & 60.88 & 60.87 & 80.92 & 63.35 & 63.29 & 88.16 & 73.39 & 73.62 & 92.08 & 67.36 & 67.23 & 93.97 & 70.49 & 70.28 & 96.50 & 67.59 & 67.38      \\
\textit{w}/\textit{o} C-PA & 82.31 & 61.15 & 61.13 & 80.92 & 63.46 & 63.39 & 88.26 & 73.63 & 73.85 & 92.26 & 67.60 & 67.39 & 94.17 & 70.71 & 70.38 & 96.58 & 67.95 & 68.02     \\
\textit{w}/\textit{o} C-PF & 82.09 & 61.05 & 61.04 & 80.91 & 63.48 & 63.41 & 87.91 & 72.59 & 72.87 & 91.96 & 67.46 & 67.31 & 93.91 & 70.40 & 70.18 & 96.39 & 67.74 & 67.83     \\
\bottomrule
\end{tabular*}
\end{table*}

\begin{table*}[t]
\small
\centering
\caption{Offline performance(\%) of FARM in terms of AUC, UAUC and GAUC in live-streaming services on Kuaishou Lite. Improve $\uparrow$ represents the improvements(\%) of FARM over PLE (base) model.}
\label{mainexperiment_2}
\setlength\tabcolsep{2.0pt}
\begin{tabular*}{1 \textwidth}
{@{\extracolsep{\fill}}@{}lcccccccccccccccccc@{}}
\toprule
\multirow{2.5}{*}{\textbf{Variants}} & \multicolumn{3}{c}{\textbf{Click}}  & \multicolumn{3}{c}{\textbf{Effective-view}} & \multicolumn{3}{c}{\textbf{Long-view}} & \multicolumn{3}{c}{\textbf{Like}} & \multicolumn{3}{c}{\textbf{Comment}} & \multicolumn{3}{c}{\textbf{Gift}}
\\ \cmidrule(r){2-4} \cmidrule(r){5-7} \cmidrule(r){8-10} \cmidrule(r){11-13} \cmidrule(r){14-16} \cmidrule(r){17-19}& AUC & UAUC & GAUC & AUC & UAUC & GAUC & AUC & UAUC & GAUC & AUC & UAUC & GAUC & AUC & UAUC & GAUC & AUC & UAUC & GAUC\\
\midrule
PLE (Base) & 82.09 & 59.05 & 59.35 & 80.21 & 61.55 & 61.54 & 86.62 & 70.32 & 70.39 & 91.92 & 65.47 & 65.42 & 94.33 & 67.64 & 67.59 & 96.77 & 64.06 & 63.40       \\
\midrule
FARM & 82.62 & 59.75 & 60.01 & 80.39 & 61.89 & 61.90 & 87.23 & 72.05 & 71.79 & 92.41 & 66.09 & 66.02 & 94.72 & 68.68 & 68.57 & 97.05 & 65.27 & 64.85       \\
Improve $\uparrow$ & \textbf{+0.53} & \textbf{+0.70} & \textbf{+0.66} & \textbf{+0.18} & \textbf{+0.34} & \textbf{+0.36} & \textbf{+0.61} & \textbf{+1.73} & \textbf{+1.40} & \textbf{+0.49} & \textbf{+0.62} & \textbf{+0.60} & \textbf{+0.39} & \textbf{+1.04} & \textbf{+0.98} & \textbf{+0.28} & \textbf{+1.21} & \textbf{+1.45}       \\
\midrule
\textit{w}/\textit{o} V-FA & 82.59 & 59.71 & 59.99 & 80.37 & 61.79 & 61.80 & 87.22 & 71.93 & 71.74 & 92.35 & 66.08 & 65.77 & 94.68 & 68.42 & 68.27 & 96.96 & 65.19 & 64.81       \\
\textit{w}/\textit{o} L-FA & 82.49 & 59.54 & 59.85 & 80.33 & 61.71 & 61.72 & 87.13 & 71.82 & 71.61 & 92.22 & 66.03 & 65.84 & 94.55 & 68.44 & 68.33 & 96.93 & 65.13 & 64.75      \\
\textit{w}/\textit{o} C-PA & 82.60 & 59.72 & 59.99 & 80.35 & 61.84 & 61.85 & 87.21 & 72.02 & 71.78 & 92.38 & 66.07 & 65.95 & 94.70 & 68.39 & 68.30 & 97.01 & 65.22 & 64.82     \\
\textit{w}/\textit{o} C-PF & 82.30 & 59.61 & 59.87 & 80.32 & 61.83 & 61.85 & 86.86 & 71.10 & 70.96 & 92.05 & 66.01 & 65.79 & 94.48 & 68.03 & 68.00 & 96.85 & 65.08 & 64.78     \\
\bottomrule
\end{tabular*}
\end{table*}

\subsubsection{\textbf{Base Models and Evaluation Metrics}}
The multi-gate mixture-of-experts model plays a crucial role in estimating the probabilities of various interactions within industry ranking models. Moreover, it has multiple configuration options, \textit{e.g.}, MMoE~\cite{mmoe}, CGC~\cite{ple}, PLE~\cite{ple}, AdaTT~\cite{adatt}, etc. In this paper, we choose \textbf{PLE}, a representative multi-task learning method as base model to evaluate the effectiveness of our FARM.

Following previous works~\cite{mmbee,moment_cross,qarm}, we evaluate the performance of our FARM and base model with widely adopted metrics: AUC, UAUC and GAUC. AUC represents the probability that the score of a positive user-item sample pair is higher than that of a negative user-item sample pair. UAUC calculates the average AUC value for different users. Moreover, GAUC is the the weighted version of UAUC, which integrates different users interaction ratios. They are formulated as follows: 
\begin{equation}
\begin{split}
\text{UAUC}=\frac{1}{N}\sum_{i=1}^{N}\text{AUC}_i,\quad
\text{GAUC}=\sum_{i=1}^{N}\frac{sample_i}{all~sample}\text{AUC}_i,
\end{split}
\label{}
\end{equation}
where $N$ denotes the number of users in the testing set. UAUC and GAUC mitigate the bias among users, enabling a more precise and equitable evaluation of the model’s performance.

\subsubsection{\textbf{Implementation Details}}
We implement our method with TensorFlow on Tesla T4 GPU. We optimize all models using the Adam~\cite{adam} optimizer, with the batch size is fixed at 5000. For our FARM, the embedding dimension $D_1$ is set to 92, $D_2$ is set to 64. The sequence length of ID-based features and Side Info-based features in both domains are set to 50. The number of MLP layer is chosen from $\{2,3,4,5,6\}$. The cut-off frequency $c$ is chosen from $\{1,3,5,7,9\}$. The hyper-parameter $\lambda$ is chosen from $\{0.1,0.3,0.5,0.7,0.9\}$. We set the temperature coefficient $\tau$ to 0.07. Note that we report the average performance of all models on testing set over hours.

\subsection{Offline Performance (RQ1)}
In this section, we compare the offline performance of our FARM against base model PLE on live-streaming services, including Kuaishou and Kuaishou Lite in Table~\ref{mainexperiment_1} and \ref{mainexperiment_2}, respectively. Specifically, our online live-streaming services need to handle billions of user requests daily, and even a 0.10\% improvement in offline AUC, GAUC and GAUC evaluations is substantial enough to yield measurable performance gains in the online services. We also develop four distinct model variants to establish a comprehensive ablation analysis with our original model, which are outlined as follows: 
\begin{itemize}
    \item \textit{w}/\textit{o} V-FA: This variant excludes the video-domain frequency aware module in Section~\ref{fq_modeling}.
    \item \textit{w}/\textit{o} L-FA: This variant involves the removal of the live-domain frequency aware module in Section~\ref{fq_modeling}.
    \item \textit{w}/\textit{o} C-PA: In this variant, we remove the cross-domain preference align module in Section ~\ref{cl_modeling}.
    \item \textit{w}/\textit{o} C-PF: This variant removes the cross-domain preference fuse module in Section~\ref{fuse_modeling}.
\end{itemize}
From the results shown in Table~\ref{mainexperiment_1} and \ref{mainexperiment_2}, we find that the removal of intra-domain (\textit{i.e.}, short-video domain and live-streaming domain) frequency aware module leads to performance degradation. This observation validates the effectiveness of perceiving user's sparse yet valuable behaviors (\textit{i.e.}, high-frequency information) to accurately describe user's personalized preference using the DFT. Meanwhile, we also observe that FARM significantly outperforms the \textit{w}/\textit{o} C-PA variant, which demonstrate the crucial role of cross-domain preference align module to narrow the gap between user preference representations in both domains. Another observation is that the performance decreases the most after the removal of cross-domain preference fuse module, which further indicates the necessary to fuse user preference in both domains with a serious well-designed attention mechanisms, thus achieving the preference transfer across domains and fully modeling user preference. Overall, all modules we proposed in our FARM can bring performance improvements, and the unification of all modules achieves an average improvement of 0.42\%, 1.06\% and 1.05\% with respect to AUC, UAUC and GAUC metrics over Kuaishou and Kuaishou Lite.

\subsection{\textbf{Frequency-Aware Analysis (RQ2)}}
To better understand the ability of the intra-domain frequency-aware module to enable our model to perceive user's sparse yet valuable behaviors, we construct a case study to compare FARM with \textit{w}/\textit{o} V-FA \& L-FA variant (\textit{i.e.}, the removal of both video-domain and live-domain frequency-aware modules) in Table~\ref{case_study}. Specifically, we randomly select an low-active user who rarely gives like, comment and gift behaviors to live-streaming authors, which means this user's valuable behaviors are sparse and hard to perceive. Subsequently, we randomly select a candidate author who is given like and gift by this user (\textit{i.e.}, $y^{ltr}_t=1$, $y^{gtr}_t=1$), and these behaviors are successfully predicted by our FARM (\textit{i.e.}, $\hat{y}^{ltr}_t=0.91$, $\hat{y}^{gtr}_t=0.88$). On the contrary, \textit{w}/\textit{o} V-FA \& L-FA variant cannot make accurate prediction for these sparse yet valuable behaviors (\textit{i.e.}, $\hat{y}^{ltr}_t=0.41$, $\hat{y}^{gtr}_t=0.36$). We can also draw similar observations from the perspective of the trainable parameters (\textit{i.e.}, $\beta$, $\delta$ in Eq.~(\ref{eq_trainable})), which control the weights of high-frequency information in both domains. From the results shown in Table~\ref{case_study}, we find that our FARM tends to assign higher weights to high-frequency information, thus perceiving user's sparse yet valuable behaviors.

\begin{table}[t]
\centering
\small
\caption{A case study of a user in live-streaming services.}
\setlength{\tabcolsep}{2pt}
\label{case_study}
\begin{tabular*}{0.45 \textwidth}{@{\extracolsep{\fill}}@{}c|ccc@{}}
\toprule
\multirow{2}{*}{\textbf{Variants}} & \multirow{2}{*}{\textbf{Candidate author}}  &  \textbf{Predicted}  & \textbf{Trainable}    \\
&  &  \textbf{probabilities}  &  \textbf{parameters}   \\
\midrule
\multirow{3}{*}{FARM} & \multirow{6}{*}{\tikz[baseline={(current bounding box.center)}]{\node[draw=red, line width=0.25mm, inner sep=1.9] {\includegraphics[width=1.18cm]{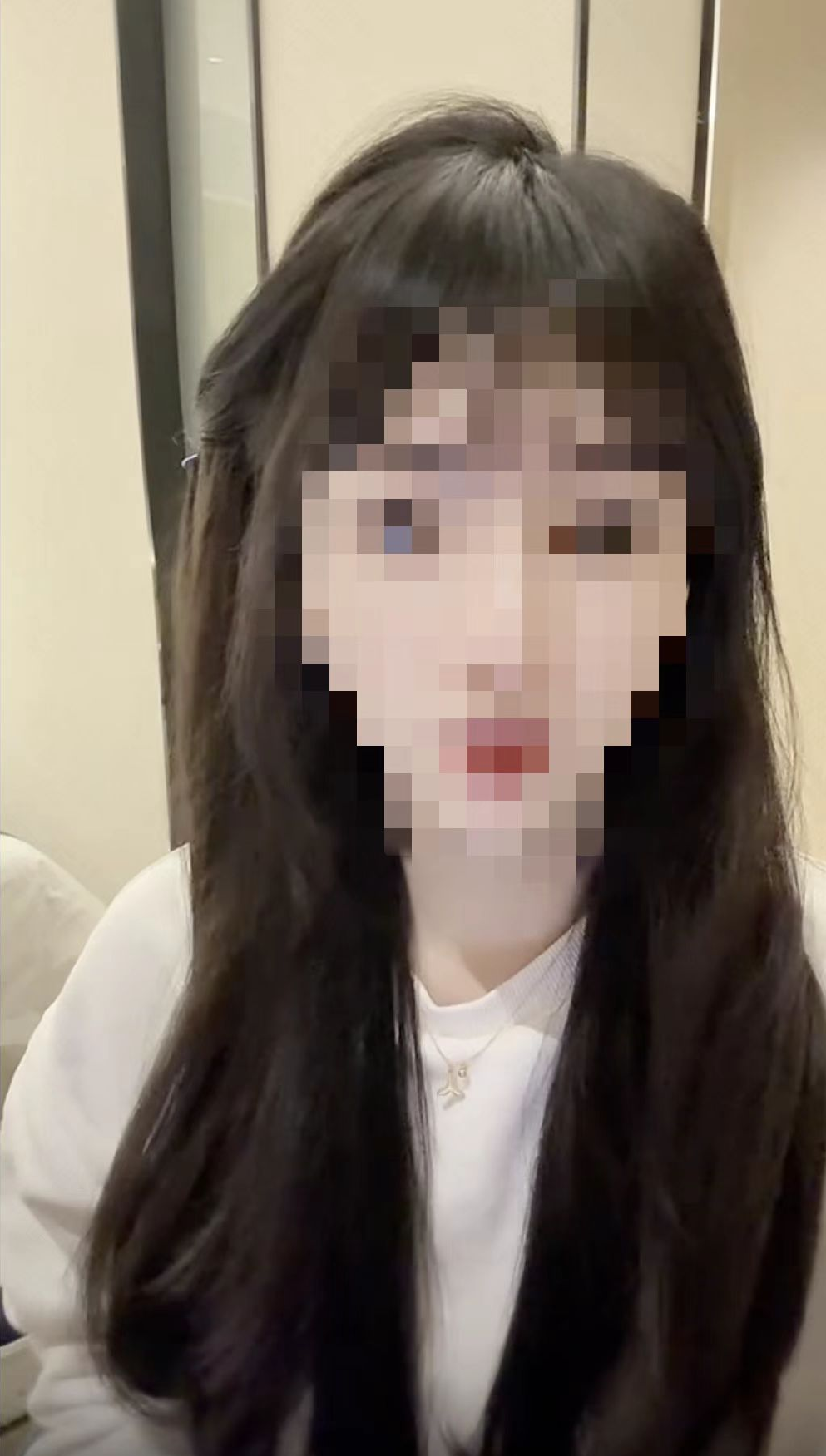}};}}  & \multirow{3}{*}{\makecell{$\hat{y}^{ltr}_t=0.91 \uparrow$\\$\hat{y}^{gtr}_t=0.88 \uparrow$}}   &  \multirow{3}{*}{\makecell{$\beta=0.89 \uparrow$\\$\delta=0.93 \uparrow$}} \\
&  &  &  \\
&  &  &   \\
\cmidrule{1-1}\cmidrule{3-4}
\multirow{3}{*}{\textit{w}/\textit{o} V-FA \& L-FA} &   &  \multirow{3}{*}{\makecell{$\hat{y}^{ltr}_t=0.41 \downarrow$\\$\hat{y}^{gtr}_t=0.36 \downarrow$}} &  \multirow{3}{*}{---}    \\
&  &  &   \\
&  &  &   \\

\bottomrule
\end{tabular*}
\end{table}

\begin{figure}[t]
\setlength{\abovecaptionskip}{0.cm}
	\begin{center}
        \subfigure
        {\begin{minipage}[b]{.47\linewidth}
        \centering
        \includegraphics[scale=0.31]{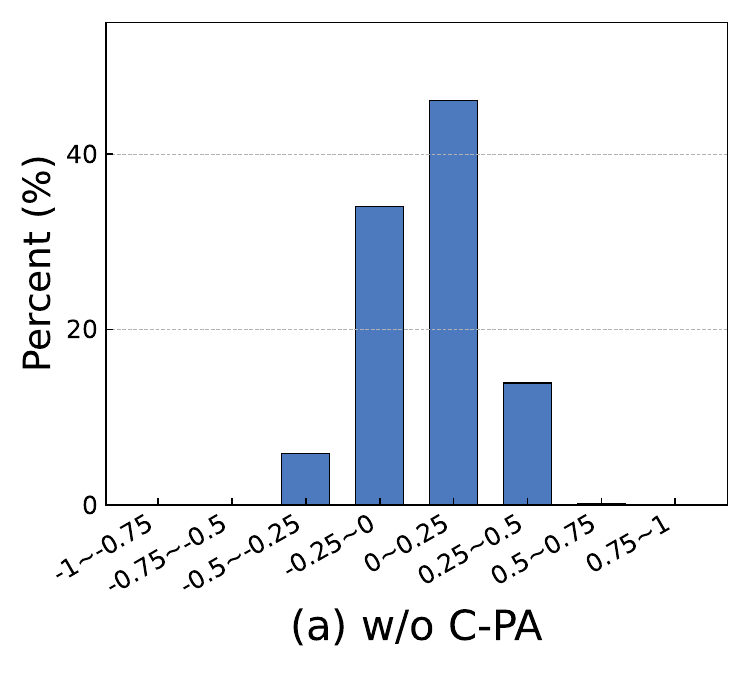}
        \end{minipage}}
        \subfigure
        {\begin{minipage}[b]{.47\linewidth}
        \centering
        \includegraphics[scale=0.31]{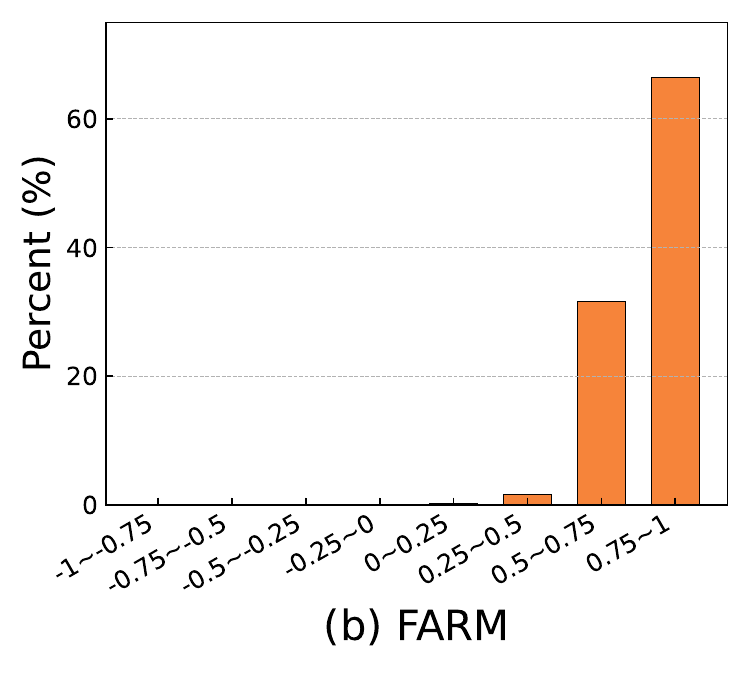}
        \end{minipage}}
        
	\caption{A statistical result of the \textit{cosine} similarity with respect to (a) \textit{w}/\textit{o} C-PA variant, and (b) FARM.}
	\label{case_cl}
	\end{center}
\vspace{-0.9em}
\end{figure}

\subsection{\textbf{Preference Align Analysis (RQ3)}}
To explore the capability of our cross-domain preference align module to narrow the gap between user preference representation in both domains, we conduct a statistical analysis to compare FARM with \textit{w}/\textit{o} C-PA variant in Figure~\ref{case_cl}. Specifically, we randomly sample 10000 users and calculate their \textit{cosine} similarity (ranging from -1 to 1) of the preference representation $\bm{h}^{video}$ in the short-video domain and $\bm{h}^{live}$ in the live-streaming domain (refer to Eq.~(\ref{eq_mlp})). From the results shown in Figure~\ref{case_cl}, we observe that the \textit{cosine} similarity of user preference representations in both domains of our FARM is much higher than that of the \textit{w}/\textit{o} C-PA variant, which indicates that our proposed cross-domain preference align module can effectively align user preference in both domains.

\begin{figure}[t!]
\begin{center}
\includegraphics[width=8.5cm]{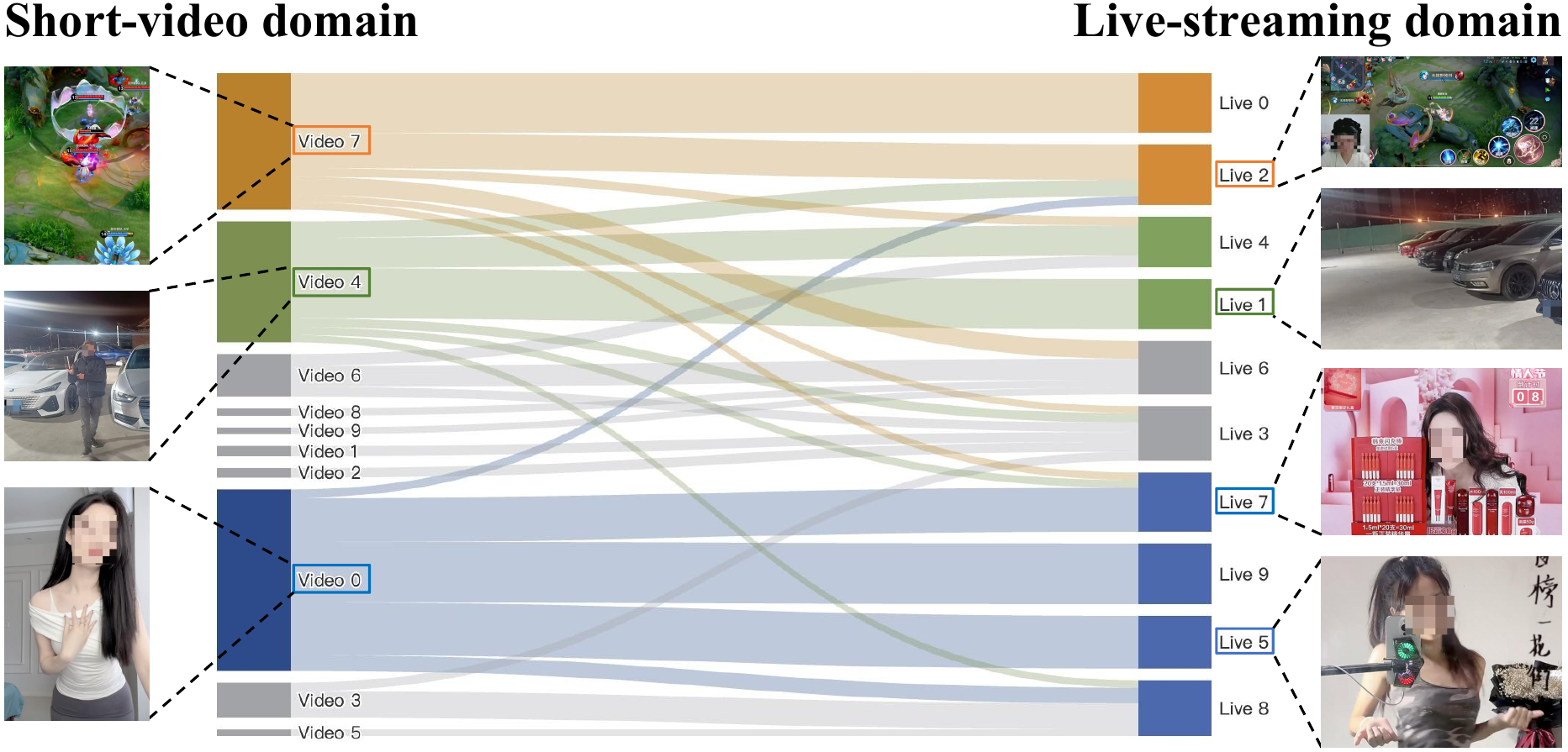}
\caption{Visualization of the fused user preference.}
\vspace{-0.9em}
\label{case_cross}
\end{center}
\end{figure}

\subsection{\textbf{Preference Fuse Analysis (RQ4)}}
To verify the effectiveness of the cross-domain preference fuse module to transfer user preference across the short-video and live-streaming domains, we conduct a visualization study to show the weights of the fused user preference $\hat{\mathbf{V}}^{cross}$ in Eq.~(\ref{eq_cross_attention}). Specifically, we randomly sample a user and visualize the weights of this user's fused preference in Figure~\ref{case_cross}, where the left denotes the author ID of user's historical viewing in the short-video domain, and the right denotes the author ID in the live-streaming domain. From the visualization results shown in Figure~\ref{case_cross}, we find that FARM can well fuse users’ similar preferences in both domains. For example, our FARM tends to assign higher weights between Video 7 and Live 2, both of which are game-related content. Moreover, we also observe that FARM can stimulate user's potential preference in the live-streaming domain by transferring preference in the short-video domain. Since our model assigns higher weights to Video 0 and Live 7, which stimulates user's desire to purchase in the live-streaming domain and brings revenue to our platform. Besides, we find that FARM is able to transfer user preference in the short-video domain to mitigate the data-sparsity issue in the live-streaming domain. For instance, Video 4 and Live 1 are assigned higher weights, enabling our model to recommend second-hand car transaction live-streaming, which is a minority category in our live-streaming services and it is difficult to recommend such live-streaming without user preference from the short-video domain.

\begin{table}[t]
\centering
\small
\caption{Improvements of FARM over PLE (base) model with different user groups on Kuaishou.}
\setlength{\tabcolsep}{9.5pt}{
\begin{tabular}{ccccc}
\toprule
\multirow{2.5}{*}{\textbf{Metrics}} & \multicolumn{4}{c}{\textbf{Active User Groups}} \\
\cmidrule{2-5} & Low  & Mid & High & Full \\
\midrule
Click & +0.65\% & +0.53\% & +0.47\% & +0.40\%  \\
Long-view & +0.65\% & +0.61\% & +0.57\% & +0.55\%    \\
Like & +0.74\% & +0.63\% & +0.49\% & +0.31\%   \\
Comment & +0.52\% & +0.51\% & +0.47\%  & +0.33\%    \\
Gift & +0.32\% & +0.41\% & +0.29\% & +0.14\%   \\
\bottomrule
\end{tabular}
}
\label{user_groups}
\end{table}

\subsection{\textbf{Study of Different User Groups (RQ5)}}
We further explore the ability of our FARM to mitigate the data-sparsity issue from the perspective of different user groups. Specifically, we divide users on our platform into four groups (\textit{i.e.}, Low, Mid, High and Full Active) based on their active levels. We report the results of AUC in Table~\ref{user_groups}, while the results of UAUC and GAUC with similar observations are omitted to save space. We find that our FARM delivers significant improvements across all user groups, demonstrating the superiority of our proposed modules and providing a better experience for all users. Moreover, we also observe that the performance improvements brought by FARM for Low/Mid active (sparse behaviors) users are much more significant than that for High/Full active users. Such observation further indicates the effectiveness of our FARM to alleviate the data-sparsity issue in live-streaming services on Kuaishou.

\begin{table*}[t!]
\centering
\small
\setlength{\tabcolsep}{8pt}{
\caption{Online A/B testing performance of our FARM on live-streaming services at Kuaishou and Kuaishou Lite.}
\begin{tabular}{cccccccc}
\toprule
\multirow{2.3}{*}{\textbf{Applications}}  & \multicolumn{4}{c}{\textbf{Core Metrics}}  & \multicolumn{3}{c}{\textbf{Interaction Metrics}}   \\
\cmidrule(r){2-5}  \cmidrule(r){6-8}
 & Click   & Watch Time   & Gift Count  &Exposure & Like  & Comment  & Follow \\
\midrule
Kuaishou  &+0.41\% &+0.20\% &+0.77\% &+0.41\% &+1.57\% &+0.59\% &+0.96\%  \\
\midrule
Kuaishou Lite  &+0.35\% &+0.07\% &+3.77\% &+0.64\% &+3.03\% &+0.18\% & +1.75\% \\
\bottomrule
\label{ab_test}
\end{tabular}
}
\vspace{-0.4cm}
\end{table*}

\subsection{Online A/B Testing (RQ6)}
In this section, we conduct rigorous online A/B testing in our live-streaming services on two applications, including Kuaishou and Kuaishou Lite, spanning from 2025/01/03 to 2025/01/09. We evaluate the performance of FARM based on the core metrics (\textit{e.g.}, Click, Gift Count, etc) and interaction metrics (\textit{e.g.}, Like, Comment, etc). The results of online A/B testing are shown in Table~\ref{ab_test}. We can see that FARM achieves significant improvements of \textbf{+0.77\%}/\textbf{+3.77\%} in Gift Count, \textbf{+1.57\%}/\textbf{+3.03\%} in Like and \textbf{+0.96\%}/\textbf{+1.75\%} in Follow, which demonstrates the practical applicability and effectiveness of our FARM in live-streaming services, bringing considerable revenue growth to the platform.

\section{Related Work}

\textbf{Live-streaming recommendation}~\cite{contentctr,kuaihl,mmbee,moment_cross} has been widely explored in many online platforms, such as Kuaishou and TikTok. 
Compared to short-video recommendation~\cite{m3csr,cvrdd}, live-streaming recommendation is more challenging due to its real-time and interactive nature.
One representative early study LiveRec~\cite{liverec} proposes a self-attentive model to personalize item ranking by exploiting historical interactions and current availability. ADARM~\cite{adarm} models matching patterns by capturing changes in user and author preferences. 
Some studies also consider the live-streaming multi-modal information. Specifically, ContentCTR~\cite{contentctr} and KuaiHL~\cite{kuaihl} adopt the multi-modal transformer, which combines real-time visual, acoustic and textual information to identify the highlight moment of authors. Besides, some studies further utilize GNN to capture the dynamic relationship between users and authors.  
For instance, MMBee~\cite{mmbee} constructs user-to-author and author-to-author graphs to combine adjacency and multi-hop relationships to guide interest expansion. 
However, all these methods above only model user preference within the live-streaming domains, leading to the serious data-sparsity issue. On the contrary, FARM considers to transfer user preference across the short-video and live-streaming domains, thus alleviating the data-sparsity issue.

\textbf{Cross-domain recommendation} (CDR)~\cite{cdrib, cdrnp, dmcdr} aims at leveraging user interactions in the source domain to improve the recommendation performance in the target domain, which is a valid direction to solve the cold-start and the data-sparsity issues in industry. 
As our FARM aim to solve the latter issue, we only focus on CDR methods that alleviate the data-sparsity issue. 
Specifically, some methods focus on capturing the domain-shared information across domains. PEPNet~\cite{pepnet} designs shared and separate parameter networks to capture the domain-shared information and achieve the balance of personalization tasks in each domain. 
CDAnet~\cite{cdanet} develops a translation network to explicitly transfer domain-shared knowledge across domains.
The others focus on combating negative transfer. For instance, DDCDR~\cite{ddcdr} leverages the disentangling distillation of cross-domain teacher models to achieve multi-domain knowledge sharing of student models.  
ADSNet~\cite{adsnet} introduces a gain evaluation strategy to calculate information gain, which is used for the rejection of noisy samples.
More recently, some studies apply CDR in live-streaming recommendation. Moment\&Cross~\cite{moment_cross} follows the search-based~\cite{sim} interest modeling idea to extract user preference in both the short-video and live-streaming domains. LCN~\cite{lcn} achieves noise filtering through a lifelong attention pyramid containing three cascading attention levels. Compared to them, FARM has important design differences that can perceive user's sparse yet valuable behaviors.

\textbf{Fourier transform}~\cite{fft} is a mathematical tool that converts time-domain signals into frequency-domain signals, and has recently been widely employed in recommendation systems~\cite{farima,kda,bsarec,fmlp-rec,end4rec} due to its advantages in capturing periodic patterns and reducing dimensionality.  
These methods have two main concerns, one is the periodic high-order modeling of user sequences. Specifically, FARIMA~\cite{farima} develops the Fourier-assisted Auto-Regressive approach to analyze the periodicity and randomness of user interactions. KDA~\cite{kda} models the temporal drift of relations between users and items via Fourier transform with learnable frequency domain embedding. BSARec~\cite{bsarec} leverages the Fourier transform to integrate low and high frequency information to mitigate oversmoothing and address the limitations of self-attention. Some studies also focus on denoising by filtering in frequency domain. For instance, FMLP-Rec~\cite{fmlp-rec} applies fast Fourier transform to sequence representation and achieves noise attenuation through learnable filters. END4Rec~\cite{end4rec} obtains the soft noise signal and the filtered signal through chunked diagonal mechanism and token sparsity in the frequency domain. Different from prior methods, we propose a frequency-aware model in this paper, which innovatively adopts the Discrete Fourier Transform (DFT) theory into cross-domain live-streaming recommendation to perceive user's sparse yet valuable behaviors in both domains.

\section{Conclusion}
In this paper, we propose a novel and practical frequency-aware model named FARM for cross-domain live-streaming recommendation, which alleviate the data-sparsity issue from the perspective of user behaviors and exposure content. We develop the intra-domain frequency-aware module to enable our model to perceive user's sparse yet valuable behaviors in both the short-video and live-streaming domains. Additionally, we apply the preference align before fuse strategy to transfer user preference across the short-video and live-streaming domains. Extensive offline experiments and online A/B testing further verify the effectiveness and superiority of FARM in live-streaming services on Kuaishou.

\balance
\bibliographystyle{ACM-Reference-Format}
\bibliography{sample-base-extend.bib}


\begin{thebibliography}{43}


\ifx \showCODEN    \undefined \def \showCODEN     #1{\unskip}     \fi
\ifx \showDOI      \undefined \def \showDOI       #1{#1}\fi
\ifx \showISBNx    \undefined \def \showISBNx     #1{\unskip}     \fi
\ifx \showISBNxiii \undefined \def \showISBNxiii  #1{\unskip}     \fi
\ifx \showISSN     \undefined \def \showISSN      #1{\unskip}     \fi
\ifx \showLCCN     \undefined \def \showLCCN      #1{\unskip}     \fi
\ifx \shownote     \undefined \def \shownote      #1{#1}          \fi
\ifx \showarticletitle \undefined \def \showarticletitle #1{#1}   \fi
\ifx \showURL      \undefined \def \showURL       {\relax}        \fi
\providecommand\bibfield[2]{#2}
\providecommand\bibinfo[2]{#2}
\providecommand\natexlab[1]{#1}
\providecommand\showeprint[2][]{arXiv:#2}

\bibitem[An et~al\mbox{.}(2024)]%
        {ddcdr}
\bibfield{author}{\bibinfo{person}{Zhicheng An}, \bibinfo{person}{Zhexu Gu}, \bibinfo{person}{Li Yu}, \bibinfo{person}{Ke Tu}, \bibinfo{person}{Zhengwei Wu}, \bibinfo{person}{Binbin Hu}, \bibinfo{person}{Zhiqiang Zhang}, \bibinfo{person}{Lihong Gu}, {and} \bibinfo{person}{Jinjie Gu}.} \bibinfo{year}{2024}\natexlab{}.
\newblock \showarticletitle{DDCDR: A Disentangle-based Distillation Framework for Cross-Domain Recommendation}. In \bibinfo{booktitle}{\emph{Proceedings of the 30th ACM SIGKDD Conference on Knowledge Discovery and Data Mining}}. \bibinfo{pages}{4764--4773}.
\newblock


\bibitem[Cao et~al\mbox{.}(2022a)]%
        {cdrib}
\bibfield{author}{\bibinfo{person}{Jiangxia Cao}, \bibinfo{person}{Jiawei Sheng}, \bibinfo{person}{Xin Cong}, \bibinfo{person}{Tingwen Liu}, {and} \bibinfo{person}{Bin Wang}.} \bibinfo{year}{2022}\natexlab{a}.
\newblock \showarticletitle{Cross-domain recommendation to cold-start users via variational information bottleneck}. In \bibinfo{booktitle}{\emph{2022 IEEE 38th International Conference on Data Engineering (ICDE)}}.
\newblock


\bibitem[Cao et~al\mbox{.}(2024)]%
        {moment_cross}
\bibfield{author}{\bibinfo{person}{Jiangxia Cao}, \bibinfo{person}{Shen Wang}, \bibinfo{person}{Yue Li}, \bibinfo{person}{Shenghui Wang}, \bibinfo{person}{Jian Tang}, \bibinfo{person}{Shiyao Wang}, \bibinfo{person}{Shuang Yang}, \bibinfo{person}{Zhaojie Liu}, {and} \bibinfo{person}{Guorui Zhou}.} \bibinfo{year}{2024}\natexlab{}.
\newblock \showarticletitle{Moment\&Cross: Next-Generation Real-Time Cross-Domain CTR Prediction for Live-Streaming Recommendation at Kuaishou}.
\newblock \bibinfo{journal}{\emph{arXiv preprint arXiv:2408.05709}} (\bibinfo{year}{2024}).
\newblock


\bibitem[Cao et~al\mbox{.}(2022b)]%
        {sdim}
\bibfield{author}{\bibinfo{person}{Yue Cao}, \bibinfo{person}{Xiaojiang Zhou}, \bibinfo{person}{Jiaqi Feng}, \bibinfo{person}{Peihao Huang}, \bibinfo{person}{Yao Xiao}, \bibinfo{person}{Dayao Chen}, {and} \bibinfo{person}{Sheng Chen}.} \bibinfo{year}{2022}\natexlab{b}.
\newblock \showarticletitle{Sampling is all you need on modeling long-term user behaviors for CTR prediction}. In \bibinfo{booktitle}{\emph{Proceedings of the 31st ACM International Conference on Information \& Knowledge Management}}. \bibinfo{pages}{2974--2983}.
\newblock


\bibitem[Chang et~al\mbox{.}(2023a)]%
        {twin}
\bibfield{author}{\bibinfo{person}{Jianxin Chang}, \bibinfo{person}{Chenbin Zhang}, \bibinfo{person}{Zhiyi Fu}, \bibinfo{person}{Xiaoxue Zang}, \bibinfo{person}{Lin Guan}, \bibinfo{person}{Jing Lu}, \bibinfo{person}{Yiqun Hui}, \bibinfo{person}{Dewei Leng}, \bibinfo{person}{Yanan Niu}, \bibinfo{person}{Yang Song}, {et~al\mbox{.}}} \bibinfo{year}{2023}\natexlab{a}.
\newblock \showarticletitle{TWIN: TWo-stage interest network for lifelong user behavior modeling in CTR prediction at kuaishou}. In \bibinfo{booktitle}{\emph{Proceedings of the 29th ACM SIGKDD Conference on Knowledge Discovery and Data Mining}}. \bibinfo{pages}{3785--3794}.
\newblock


\bibitem[Chang et~al\mbox{.}(2023b)]%
        {pepnet}
\bibfield{author}{\bibinfo{person}{Jianxin Chang}, \bibinfo{person}{Chenbin Zhang}, \bibinfo{person}{Yiqun Hui}, \bibinfo{person}{Dewei Leng}, \bibinfo{person}{Yanan Niu}, \bibinfo{person}{Yang Song}, {and} \bibinfo{person}{Kun Gai}.} \bibinfo{year}{2023}\natexlab{b}.
\newblock \showarticletitle{Pepnet: Parameter and embedding personalized network for infusing with personalized prior information}. In \bibinfo{booktitle}{\emph{Proceedings of the 29th ACM SIGKDD Conference on Knowledge Discovery and Data Mining}}. \bibinfo{pages}{3795--3804}.
\newblock


\bibitem[Chen et~al\mbox{.}(2024b)]%
        {m3csr}
\bibfield{author}{\bibinfo{person}{Gaode Chen}, \bibinfo{person}{Ruina Sun}, \bibinfo{person}{Yuezihan Jiang}, \bibinfo{person}{Jiangxia Cao}, \bibinfo{person}{Qi Zhang}, \bibinfo{person}{Jingjian Lin}, \bibinfo{person}{Han Li}, \bibinfo{person}{Kun Gai}, {and} \bibinfo{person}{Xinghua Zhang}.} \bibinfo{year}{2024}\natexlab{b}.
\newblock \showarticletitle{A Multi-modal Modeling Framework for Cold-start Short-video Recommendation}. In \bibinfo{booktitle}{\emph{Proceedings of the 18th ACM Conference on Recommender Systems}}. \bibinfo{pages}{391--400}.
\newblock


\bibitem[Chen et~al\mbox{.}(2021)]%
        {eta}
\bibfield{author}{\bibinfo{person}{Qiwei Chen}, \bibinfo{person}{Changhua Pei}, \bibinfo{person}{Shanshan Lv}, \bibinfo{person}{Chao Li}, \bibinfo{person}{Junfeng Ge}, {and} \bibinfo{person}{Wenwu Ou}.} \bibinfo{year}{2021}\natexlab{}.
\newblock \showarticletitle{End-to-end user behavior retrieval in click-through rateprediction model}.
\newblock \bibinfo{journal}{\emph{arXiv preprint arXiv:2108.04468}} (\bibinfo{year}{2021}).
\newblock


\bibitem[Chen et~al\mbox{.}(2024a)]%
        {cdanet}
\bibfield{author}{\bibinfo{person}{Xu Chen}, \bibinfo{person}{Zida Cheng}, \bibinfo{person}{Jiangchao Yao}, \bibinfo{person}{Chen Ju}, \bibinfo{person}{Weilin Huang}, \bibinfo{person}{Jinsong Lan}, \bibinfo{person}{Xiaoyi Zeng}, {and} \bibinfo{person}{Shuai Xiao}.} \bibinfo{year}{2024}\natexlab{a}.
\newblock \showarticletitle{Enhancing cross-domain click-through rate prediction via explicit feature augmentation}. In \bibinfo{booktitle}{\emph{Companion Proceedings of the ACM on Web Conference 2024}}. \bibinfo{pages}{423--432}.
\newblock


\bibitem[Deng et~al\mbox{.}(2023)]%
        {contentctr}
\bibfield{author}{\bibinfo{person}{Jiaxin Deng}, \bibinfo{person}{Dong Shen}, \bibinfo{person}{Shiyao Wang}, \bibinfo{person}{Xiangyu Wu}, \bibinfo{person}{Fan Yang}, \bibinfo{person}{Guorui Zhou}, {and} \bibinfo{person}{Gaofeng Meng}.} \bibinfo{year}{2023}\natexlab{}.
\newblock \showarticletitle{ContentCTR: Frame-level Live Streaming Click-Through Rate Prediction with Multimodal Transformer}.
\newblock \bibinfo{journal}{\emph{arXiv preprint arXiv:2306.14392}} (\bibinfo{year}{2023}).
\newblock


\bibitem[Deng et~al\mbox{.}(2024a)]%
        {kuaihl}
\bibfield{author}{\bibinfo{person}{Jiaxin Deng}, \bibinfo{person}{Shiyao Wang}, \bibinfo{person}{Dong Shen}, \bibinfo{person}{Liqin Zhao}, \bibinfo{person}{Fan Yang}, \bibinfo{person}{Guorui Zhou}, {and} \bibinfo{person}{Gaofeng Meng}.} \bibinfo{year}{2024}\natexlab{a}.
\newblock \showarticletitle{A Multimodal Transformer for Live Streaming Highlight Prediction}. In \bibinfo{booktitle}{\emph{2024 IEEE International Conference on Multimedia and Expo (ICME)}}.
\newblock


\bibitem[Deng et~al\mbox{.}(2024b)]%
        {mmbee}
\bibfield{author}{\bibinfo{person}{Jiaxin Deng}, \bibinfo{person}{Shiyao Wang}, \bibinfo{person}{Yuchen Wang}, \bibinfo{person}{Jiansong Qi}, \bibinfo{person}{Liqin Zhao}, \bibinfo{person}{Guorui Zhou}, {and} \bibinfo{person}{Gaofeng Meng}.} \bibinfo{year}{2024}\natexlab{b}.
\newblock \showarticletitle{MMBee: Live Streaming Gift-Sending Recommendations via Multi-Modal Fusion and Behaviour Expansion}. In \bibinfo{booktitle}{\emph{Proceedings of the 30th ACM SIGKDD Conference on Knowledge Discovery and Data Mining}}. \bibinfo{pages}{4896--4905}.
\newblock


\bibitem[Han et~al\mbox{.}(2024)]%
        {end4rec}
\bibfield{author}{\bibinfo{person}{Yongqiang Han}, \bibinfo{person}{Hao Wang}, \bibinfo{person}{Kefan Wang}, \bibinfo{person}{Likang Wu}, \bibinfo{person}{Zhi Li}, \bibinfo{person}{Wei Guo}, \bibinfo{person}{Yong Liu}, \bibinfo{person}{Defu Lian}, {and} \bibinfo{person}{Enhong Chen}.} \bibinfo{year}{2024}\natexlab{}.
\newblock \showarticletitle{END4Rec: Efficient Noise-Decoupling for Multi-Behavior Sequential Recommendation}.
\newblock \bibinfo{journal}{\emph{arXiv preprint arXiv:2403.17603}} (\bibinfo{year}{2024}).
\newblock


\bibitem[Hou et~al\mbox{.}(2024)]%
        {lcn}
\bibfield{author}{\bibinfo{person}{Ruijie Hou}, \bibinfo{person}{Zhaoyang Yang}, \bibinfo{person}{Yu Ming}, \bibinfo{person}{Hongyu Lu}, \bibinfo{person}{Zhuobin Zheng}, \bibinfo{person}{Yu Chen}, \bibinfo{person}{Qinsong Zeng}, {and} \bibinfo{person}{Ming Chen}.} \bibinfo{year}{2024}\natexlab{}.
\newblock \showarticletitle{Cross-Domain LifeLong Sequential Modeling for Online Click-Through Rate Prediction}. In \bibinfo{booktitle}{\emph{Proceedings of the 30th ACM SIGKDD Conference on Knowledge Discovery and Data Mining}}. \bibinfo{pages}{5116--5125}.
\newblock


\bibitem[Kingma and Ba(2014)]%
        {adam}
\bibfield{author}{\bibinfo{person}{Diederik~P Kingma} {and} \bibinfo{person}{Jimmy Ba}.} \bibinfo{year}{2014}\natexlab{}.
\newblock \showarticletitle{Adam: A method for stochastic optimization}.
\newblock \bibinfo{journal}{\emph{arXiv preprint arXiv:1412.6980}} (\bibinfo{year}{2014}).
\newblock


\bibitem[Lei et~al\mbox{.}(2021)]%
        {semi}
\bibfield{author}{\bibinfo{person}{Chenyi Lei}, \bibinfo{person}{Yong Liu}, \bibinfo{person}{Lingzi Zhang}, \bibinfo{person}{Guoxin Wang}, \bibinfo{person}{Haihong Tang}, \bibinfo{person}{Houqiang Li}, {and} \bibinfo{person}{Chunyan Miao}.} \bibinfo{year}{2021}\natexlab{}.
\newblock \showarticletitle{Semi: A sequential multi-modal information transfer network for e-commerce micro-video recommendations}. In \bibinfo{booktitle}{\emph{Proceedings of the 27th ACM SIGKDD Conference on Knowledge Discovery \& Data Mining}}.
\newblock


\bibitem[Li et~al\mbox{.}(2023)]%
        {adatt}
\bibfield{author}{\bibinfo{person}{Danwei Li}, \bibinfo{person}{Zhengyu Zhang}, \bibinfo{person}{Siyang Yuan}, \bibinfo{person}{Mingze Gao}, \bibinfo{person}{Weilin Zhang}, \bibinfo{person}{Chaofei Yang}, \bibinfo{person}{Xi Liu}, {and} \bibinfo{person}{Jiyan Yang}.} \bibinfo{year}{2023}\natexlab{}.
\newblock \showarticletitle{AdaTT: Adaptive Task-to-Task Fusion Network for Multitask Learning in Recommendations}. In \bibinfo{booktitle}{\emph{Proceedings of the 29th ACM SIGKDD Conference on Knowledge Discovery and Data Mining}}. \bibinfo{pages}{4370--4379}.
\newblock


\bibitem[Li et~al\mbox{.}(2021)]%
        {abf}
\bibfield{author}{\bibinfo{person}{Junnan Li}, \bibinfo{person}{Ramprasaath Selvaraju}, \bibinfo{person}{Akhilesh Gotmare}, \bibinfo{person}{Shafiq Joty}, \bibinfo{person}{Caiming Xiong}, {and} \bibinfo{person}{Steven Chu~Hong Hoi}.} \bibinfo{year}{2021}\natexlab{}.
\newblock \showarticletitle{Align before fuse: Vision and language representation learning with momentum distillation}.
\newblock \bibinfo{journal}{\emph{Advances in neural information processing systems}}  \bibinfo{volume}{34} (\bibinfo{year}{2021}), \bibinfo{pages}{9694--9705}.
\newblock


\bibitem[Li et~al\mbox{.}(2024)]%
        {cdrnp}
\bibfield{author}{\bibinfo{person}{Xiaodong Li}, \bibinfo{person}{Jiawei Sheng}, \bibinfo{person}{Jiangxia Cao}, \bibinfo{person}{Wenyuan Zhang}, \bibinfo{person}{Quangang Li}, {and} \bibinfo{person}{Tingwen Liu}.} \bibinfo{year}{2024}\natexlab{}.
\newblock \showarticletitle{CDRNP: Cross-Domain Recommendation to Cold-Start Users via Neural Process}. In \bibinfo{booktitle}{\emph{Proceedings of the 17th ACM International Conference on Web Search and Data Mining}}. \bibinfo{pages}{378--386}.
\newblock


\bibitem[Li et~al\mbox{.}(2025)]%
        {dmcdr}
\bibfield{author}{\bibinfo{person}{Xiaodong Li}, \bibinfo{person}{Hengzhu Tang}, \bibinfo{person}{Jiawei Sheng}, \bibinfo{person}{Xinghua Zhang}, \bibinfo{person}{Li Gao}, \bibinfo{person}{Suqi Cheng}, \bibinfo{person}{Dawei Yin}, {and} \bibinfo{person}{Tingwen Liu}.} \bibinfo{year}{2025}\natexlab{}.
\newblock \showarticletitle{Exploring Preference-Guided Diffusion Model for Cross-Domain Recommendation}.
\newblock \bibinfo{journal}{\emph{arXiv preprint arXiv:2501.11671}} (\bibinfo{year}{2025}).
\newblock


\bibitem[Liu et~al\mbox{.}(2023)]%
        {pgsp}
\bibfield{author}{\bibinfo{person}{Jiahao Liu}, \bibinfo{person}{Dongsheng Li}, \bibinfo{person}{Hansu Gu}, \bibinfo{person}{Tun Lu}, \bibinfo{person}{Peng Zhang}, \bibinfo{person}{Li Shang}, {and} \bibinfo{person}{Ning Gu}.} \bibinfo{year}{2023}\natexlab{}.
\newblock \showarticletitle{Personalized graph signal processing for collaborative filtering}. In \bibinfo{booktitle}{\emph{Proceedings of the ACM Web Conference 2023}}. \bibinfo{pages}{1264--1272}.
\newblock


\bibitem[Luo et~al\mbox{.}(2024)]%
        {qarm}
\bibfield{author}{\bibinfo{person}{Xinchen Luo}, \bibinfo{person}{Jiangxia Cao}, \bibinfo{person}{Tianyu Sun}, \bibinfo{person}{Jinkai Yu}, \bibinfo{person}{Rui Huang}, \bibinfo{person}{Wei Yuan}, \bibinfo{person}{Hezheng Lin}, \bibinfo{person}{Yichen Zheng}, \bibinfo{person}{Shiyao Wang}, \bibinfo{person}{Qigen Hu}, {et~al\mbox{.}}} \bibinfo{year}{2024}\natexlab{}.
\newblock \showarticletitle{QARM: Quantitative Alignment Multi-Modal Recommendation at Kuaishou}.
\newblock \bibinfo{journal}{\emph{arXiv preprint arXiv:2411.11739}} (\bibinfo{year}{2024}).
\newblock


\bibitem[Ma et~al\mbox{.}(2018)]%
        {mmoe}
\bibfield{author}{\bibinfo{person}{Jiaqi Ma}, \bibinfo{person}{Zhe Zhao}, \bibinfo{person}{Xinyang Yi}, \bibinfo{person}{Jilin Chen}, \bibinfo{person}{Lichan Hong}, {and} \bibinfo{person}{Ed~H Chi}.} \bibinfo{year}{2018}\natexlab{}.
\newblock \showarticletitle{Modeling task relationships in multi-task learning with multi-gate mixture-of-experts}. In \bibinfo{booktitle}{\emph{Proceedings of the 24th ACM SIGKDD international conference on knowledge discovery \& data mining}}. \bibinfo{pages}{1930--1939}.
\newblock


\bibitem[Mateos et~al\mbox{.}(2019)]%
        {cdgsp}
\bibfield{author}{\bibinfo{person}{Gonzalo Mateos}, \bibinfo{person}{Santiago Segarra}, \bibinfo{person}{Antonio~G Marques}, {and} \bibinfo{person}{Alejandro Ribeiro}.} \bibinfo{year}{2019}\natexlab{}.
\newblock \showarticletitle{Connecting the dots: Identifying network structure via graph signal processing}.
\newblock \bibinfo{journal}{\emph{IEEE Signal Processing Magazine}} \bibinfo{volume}{36}, \bibinfo{number}{3} (\bibinfo{year}{2019}), \bibinfo{pages}{16--43}.
\newblock


\bibitem[Pi et~al\mbox{.}(2020)]%
        {sim}
\bibfield{author}{\bibinfo{person}{Qi Pi}, \bibinfo{person}{Guorui Zhou}, \bibinfo{person}{Yujing Zhang}, \bibinfo{person}{Zhe Wang}, \bibinfo{person}{Lejian Ren}, \bibinfo{person}{Ying Fan}, \bibinfo{person}{Xiaoqiang Zhu}, {and} \bibinfo{person}{Kun Gai}.} \bibinfo{year}{2020}\natexlab{}.
\newblock \showarticletitle{Search-based user interest modeling with lifelong sequential behavior data for click-through rate prediction}. In \bibinfo{booktitle}{\emph{Proceedings of the 29th ACM International Conference on Information \& Knowledge Management}}.
\newblock


\bibitem[Rappaz et~al\mbox{.}(2021)]%
        {liverec}
\bibfield{author}{\bibinfo{person}{J{\'e}r{\'e}mie Rappaz}, \bibinfo{person}{Julian McAuley}, {and} \bibinfo{person}{Karl Aberer}.} \bibinfo{year}{2021}\natexlab{}.
\newblock \showarticletitle{Recommendation on live-streaming platforms: Dynamic availability and repeat consumption}. In \bibinfo{booktitle}{\emph{Proceedings of the 15th ACM Conference on Recommender Systems}}. \bibinfo{pages}{390--399}.
\newblock


\bibitem[Shen et~al\mbox{.}(2021)]%
        {gf-cf}
\bibfield{author}{\bibinfo{person}{Yifei Shen}, \bibinfo{person}{Yongji Wu}, \bibinfo{person}{Yao Zhang}, \bibinfo{person}{Caihua Shan}, \bibinfo{person}{Jun Zhang}, \bibinfo{person}{B~Khaled Letaief}, {and} \bibinfo{person}{Dongsheng Li}.} \bibinfo{year}{2021}\natexlab{}.
\newblock \showarticletitle{How powerful is graph convolution for recommendation?}. In \bibinfo{booktitle}{\emph{Proceedings of the 30th ACM international conference on information \& knowledge management}}. \bibinfo{pages}{1619--1629}.
\newblock


\bibitem[Shin et~al\mbox{.}(2024)]%
        {bsarec}
\bibfield{author}{\bibinfo{person}{Yehjin Shin}, \bibinfo{person}{Jeongwhan Choi}, \bibinfo{person}{Hyowon Wi}, {and} \bibinfo{person}{Noseong Park}.} \bibinfo{year}{2024}\natexlab{}.
\newblock \showarticletitle{An attentive inductive bias for sequential recommendation beyond the self-attention}. In \bibinfo{booktitle}{\emph{Proceedings of the AAAI Conference on Artificial Intelligence}}, Vol.~\bibinfo{volume}{38}. \bibinfo{pages}{8984--8992}.
\newblock


\bibitem[Sun et~al\mbox{.}(2023)]%
        {sitn}
\bibfield{author}{\bibinfo{person}{Guoqiang Sun}, \bibinfo{person}{Yibin Shen}, \bibinfo{person}{Sijin Zhou}, \bibinfo{person}{Xiang Chen}, \bibinfo{person}{Hongyan Liu}, \bibinfo{person}{Chunming Wu}, \bibinfo{person}{Chenyi Lei}, \bibinfo{person}{Xianhui Wei}, {and} \bibinfo{person}{Fei Fang}.} \bibinfo{year}{2023}\natexlab{}.
\newblock \showarticletitle{Self-supervised interest transfer network via prototypical contrastive learning for recommendation}. In \bibinfo{booktitle}{\emph{Proceedings of the AAAI Conference on Artificial Intelligence}}, Vol.~\bibinfo{volume}{37}. \bibinfo{pages}{4614--4622}.
\newblock


\bibitem[Tang et~al\mbox{.}(2020)]%
        {ple}
\bibfield{author}{\bibinfo{person}{Hongyan Tang}, \bibinfo{person}{Junning Liu}, \bibinfo{person}{Ming Zhao}, {and} \bibinfo{person}{Xudong Gong}.} \bibinfo{year}{2020}\natexlab{}.
\newblock \showarticletitle{Progressive layered extraction (ple): A novel multi-task learning (mtl) model for personalized recommendations}. In \bibinfo{booktitle}{\emph{Proceedings of the 14th ACM Conference on Recommender Systems}}. \bibinfo{pages}{269--278}.
\newblock


\bibitem[Tang et~al\mbox{.}(2023)]%
        {cvrdd}
\bibfield{author}{\bibinfo{person}{Shisong Tang}, \bibinfo{person}{Qing Li}, \bibinfo{person}{Dingmin Wang}, \bibinfo{person}{Ci Gao}, \bibinfo{person}{Wentao Xiao}, \bibinfo{person}{Dan Zhao}, \bibinfo{person}{Yong Jiang}, \bibinfo{person}{Qian Ma}, {and} \bibinfo{person}{Aoyang Zhang}.} \bibinfo{year}{2023}\natexlab{}.
\newblock \showarticletitle{Counterfactual video recommendation for duration debiasing}. In \bibinfo{booktitle}{\emph{Proceedings of the 29th ACM SIGKDD Conference on Knowledge Discovery and Data Mining}}. \bibinfo{pages}{4894--4903}.
\newblock


\bibitem[Van~Loan(1992)]%
        {fft}
\bibfield{author}{\bibinfo{person}{Charles Van~Loan}.} \bibinfo{year}{1992}\natexlab{}.
\newblock \bibinfo{booktitle}{\emph{Computational frameworks for the fast Fourier transform}}.
\newblock \bibinfo{publisher}{SIAM}.
\newblock


\bibitem[Vaswani et~al\mbox{.}(2017)]%
        {trans}
\bibfield{author}{\bibinfo{person}{Ashish Vaswani}, \bibinfo{person}{Noam Shazeer}, \bibinfo{person}{Niki Parmar}, \bibinfo{person}{Jakob Uszkoreit}, \bibinfo{person}{Llion Jones}, \bibinfo{person}{Aidan~N Gomez}, \bibinfo{person}{{\L}ukasz Kaiser}, {and} \bibinfo{person}{Illia Polosukhin}.} \bibinfo{year}{2017}\natexlab{}.
\newblock \showarticletitle{Attention is all you need}.
\newblock \bibinfo{journal}{\emph{Advances in neural information processing systems}}  \bibinfo{volume}{30} (\bibinfo{year}{2017}).
\newblock


\bibitem[Wang et~al\mbox{.}(2020)]%
        {kda}
\bibfield{author}{\bibinfo{person}{Chenyang Wang}, \bibinfo{person}{Weizhi Ma}, \bibinfo{person}{Min Zhang}, \bibinfo{person}{Chong Chen}, \bibinfo{person}{Yiqun Liu}, {and} \bibinfo{person}{Shaoping Ma}.} \bibinfo{year}{2020}\natexlab{}.
\newblock \showarticletitle{Toward dynamic user intention: Temporal evolutionary effects of item relations in sequential recommendation}.
\newblock \bibinfo{journal}{\emph{ACM Transactions on Information Systems (TOIS)}} \bibinfo{volume}{39}, \bibinfo{number}{2} (\bibinfo{year}{2020}), \bibinfo{pages}{1--33}.
\newblock


\bibitem[Wang et~al\mbox{.}(2024)]%
        {adsnet}
\bibfield{author}{\bibinfo{person}{Ruize Wang}, \bibinfo{person}{Hui Xu}, \bibinfo{person}{Ying Cheng}, \bibinfo{person}{Qi He}, \bibinfo{person}{Xing Zhou}, \bibinfo{person}{Rui Feng}, \bibinfo{person}{Wei Xu}, \bibinfo{person}{Lei Huang}, {and} \bibinfo{person}{Jie Jiang}.} \bibinfo{year}{2024}\natexlab{}.
\newblock \showarticletitle{ADSNet: Cross-Domain LTV Prediction with an Adaptive Siamese Network in Advertising}. In \bibinfo{booktitle}{\emph{Proceedings of the 30th ACM SIGKDD Conference on Knowledge Discovery and Data Mining}}. \bibinfo{pages}{5872--5881}.
\newblock


\bibitem[Wang and Zhang(2022)]%
        {sgnn}
\bibfield{author}{\bibinfo{person}{Xiyuan Wang} {and} \bibinfo{person}{Muhan Zhang}.} \bibinfo{year}{2022}\natexlab{}.
\newblock \showarticletitle{How powerful are spectral graph neural networks}. In \bibinfo{booktitle}{\emph{International conference on machine learning}}. PMLR, \bibinfo{pages}{23341--23362}.
\newblock


\bibitem[Xia et~al\mbox{.}(2024)]%
        {higsp}
\bibfield{author}{\bibinfo{person}{Jiafeng Xia}, \bibinfo{person}{Dongsheng Li}, \bibinfo{person}{Hansu Gu}, \bibinfo{person}{Tun Lu}, \bibinfo{person}{Peng Zhang}, \bibinfo{person}{Li Shang}, {and} \bibinfo{person}{Ning Gu}.} \bibinfo{year}{2024}\natexlab{}.
\newblock \showarticletitle{Hierarchical Graph Signal Processing for Collaborative Filtering}. In \bibinfo{booktitle}{\emph{Proceedings of the ACM on Web Conference 2024}}. \bibinfo{pages}{3229--3240}.
\newblock


\bibitem[Yi et~al\mbox{.}(2024)]%
        {fgnn}
\bibfield{author}{\bibinfo{person}{Kun Yi}, \bibinfo{person}{Qi Zhang}, \bibinfo{person}{Wei Fan}, \bibinfo{person}{Hui He}, \bibinfo{person}{Liang Hu}, \bibinfo{person}{Pengyang Wang}, \bibinfo{person}{Ning An}, \bibinfo{person}{Longbing Cao}, {and} \bibinfo{person}{Zhendong Niu}.} \bibinfo{year}{2024}\natexlab{}.
\newblock \showarticletitle{FourierGNN: Rethinking multivariate time series forecasting from a pure graph perspective}.
\newblock \bibinfo{journal}{\emph{Advances in Neural Information Processing Systems}}  \bibinfo{volume}{36} (\bibinfo{year}{2024}).
\newblock


\bibitem[Zhang et~al\mbox{.}(2021)]%
        {adarm}
\bibfield{author}{\bibinfo{person}{Shuai Zhang}, \bibinfo{person}{Hongyan Liu}, \bibinfo{person}{Jun He}, \bibinfo{person}{Sanpu Han}, {and} \bibinfo{person}{Xiaoyong Du}.} \bibinfo{year}{2021}\natexlab{}.
\newblock \showarticletitle{Deep sequential model for anchor recommendation on live streaming platforms}.
\newblock \bibinfo{journal}{\emph{Big Data Mining and Analytics}} \bibinfo{volume}{4}, \bibinfo{number}{3} (\bibinfo{year}{2021}), \bibinfo{pages}{173--182}.
\newblock


\bibitem[Zhang et~al\mbox{.}(2023)]%
        {eliverec}
\bibfield{author}{\bibinfo{person}{Yixin Zhang}, \bibinfo{person}{Yong Liu}, \bibinfo{person}{Hao Xiong}, \bibinfo{person}{Yi Liu}, \bibinfo{person}{Fuqiang Yu}, \bibinfo{person}{Wei He}, \bibinfo{person}{Yonghui Xu}, \bibinfo{person}{Lizhen Cui}, {and} \bibinfo{person}{Chunyan Miao}.} \bibinfo{year}{2023}\natexlab{}.
\newblock \showarticletitle{Cross-domain disentangled learning for e-commerce live streaming recommendation}. In \bibinfo{booktitle}{\emph{2023 IEEE 39th International Conference on Data Engineering (ICDE)}}. IEEE, \bibinfo{pages}{2955--2968}.
\newblock


\bibitem[Zhang et~al\mbox{.}(2015)]%
        {farima}
\bibfield{author}{\bibinfo{person}{Yongfeng Zhang}, \bibinfo{person}{Min Zhang}, \bibinfo{person}{Yi Zhang}, \bibinfo{person}{Guokun Lai}, \bibinfo{person}{Yiqun Liu}, \bibinfo{person}{Honghui Zhang}, {and} \bibinfo{person}{Shaoping Ma}.} \bibinfo{year}{2015}\natexlab{}.
\newblock \showarticletitle{Daily-aware personalized recommendation based on feature-level time series analysis}. In \bibinfo{booktitle}{\emph{Proceedings of the 24th international conference on world wide web}}. \bibinfo{pages}{1373--1383}.
\newblock


\bibitem[Zhou et~al\mbox{.}(2019)]%
        {dien}
\bibfield{author}{\bibinfo{person}{Guorui Zhou}, \bibinfo{person}{Na Mou}, \bibinfo{person}{Ying Fan}, \bibinfo{person}{Qi Pi}, \bibinfo{person}{Weijie Bian}, \bibinfo{person}{Chang Zhou}, \bibinfo{person}{Xiaoqiang Zhu}, {and} \bibinfo{person}{Kun Gai}.} \bibinfo{year}{2019}\natexlab{}.
\newblock \showarticletitle{Deep interest evolution network for click-through rate prediction}. In \bibinfo{booktitle}{\emph{Proceedings of the AAAI conference on artificial intelligence}}.
\newblock


\bibitem[Zhou et~al\mbox{.}(2022)]%
        {fmlp-rec}
\bibfield{author}{\bibinfo{person}{Kun Zhou}, \bibinfo{person}{Hui Yu}, \bibinfo{person}{Wayne~Xin Zhao}, {and} \bibinfo{person}{Ji-Rong Wen}.} \bibinfo{year}{2022}\natexlab{}.
\newblock \showarticletitle{Filter-enhanced MLP is all you need for sequential recommendation}. In \bibinfo{booktitle}{\emph{Proceedings of the ACM web conference 2022}}. \bibinfo{pages}{2388--2399}.
\newblock


\end{thebibliography}

\end{document}